\documentclass[aps, prd, superscriptaddress, preprintnumbers, floatfix, longbibliography,nofootinbib,twocolumn,10pt]{revtex4-2}

\usepackage[utf8]{inputenc}
\usepackage{newtxtext}
\usepackage[upint]{newtxmath}
\usepackage{microtype}
\usepackage{textcomp}
\usepackage{dsfont}
\usepackage{eucal}
\usepackage{siunitx}
\usepackage{soul}

\usepackage{graphicx,amsfonts,dsfont,amsmath,amssymb,amstext,bm}
\usepackage{float,wrapfig,psfrag}
\usepackage{xcolor}
\usepackage[colorlinks=true,urlcolor=blue,citecolor=blue,linkcolor=blue]{hyperref}
\hypersetup{breaklinks=true}
\usepackage{tcolorbox}
\usepackage[caption=false]{subfig}

\graphicspath{{Figures-paper/}} 

\newcommand{\sign}[1]{\,\mbox{sgn}\left({#1}\right)}

\newcommand{\RE}[1]{\,\mbox{Re}\left\{{#1}\right\}}
\newcommand{\IM}[1]{\,\mbox{Im}\left\{{#1}\right\}}

\newcommand{\df}[1]{\,\delta{\left(#1\right)}}

\newcommand{\tr}{\text{Tr}}

\newcommand{\vecr}{{\bf r}}

\definecolor{DarkRed}{rgb}{0.65,0,0}%
\definecolor{Green}{rgb}{0,0.3,0.3}
\definecolor{Purple}{rgb}{0.3,0,0.65}
\definecolor{Red}{rgb}{1,0,0}
\definecolor{Blue}{rgb}{0,0,0.85}
\definecolor{Magenta}{rgb}{1,0,1}

\extrafloats{100}

\begin{document}

\title{Impurity-induced Friedel oscillations in altermagnets and $p$-wave magnets}
\date{November 7, 2024}

\author{Pavlo Sukhachov}
\email{pavlo.sukhachov@ntnu.no}
\affiliation{Center for Quantum Spintronics, Department of Physics, Norwegian \\ University of Science and Technology, NO-7491 Trondheim, Norway}
\author{Jacob Linder}
\affiliation{Center for Quantum Spintronics, Department of Physics, Norwegian \\ University of Science and Technology, NO-7491 Trondheim, Norway}
\date{\today}

\begin{abstract}
We investigate the Friedel oscillations of the local density of states (LDOS) induced by a single impurity with both a spin-independent potential and an exchange coupling to the electrons in altermagnets and unconventional $p$-wave magnets. We identify features that make the Friedel oscillations and magnetization distinct from other materials with nontrivial spin texture such as Rashba metals. Because time-reversal symmetry is broken in altermagnets, both magnetic and nonmagnetic impurities lead to local magnetization with the spatial pattern that reflects the symmetry of the altermagnetic splitting. The period of the corresponding oscillations provides an alternative way to quantify the altermagnetic spin splitting and the shape of the altermagnetic bands. The LDOS pattern in $p$-wave magnets, which respect combined time-reversal and translation symmetries, is rich. It reveals anisotropy related directly to the spin splitting, but surprisingly also features LDOS oscillations with a doubled period in the proximity of the impurity. The latter effect is also observed in a Rashba metal with an exchange field and originates from the interplay of propagating and evanescent waves. The obtained results are instrumental for investigating altermagnets and unconventional $p$-wave magnets via tunneling probes.
\end{abstract}

\maketitle

\section{Introduction}
\label{sec:Introduction}

Quantum materials are at the forefront of the current theoretical and experimental studies. Topological semimetals, superconductors, and magnets are just a few examples. Recently, novel classes of magnets realizing unusual symmetry-protected spin textures have attracted significant attention.

Magnetic materials with collinear magnetic order that are invariant under combined rotations of the lattice and reversal of spins can realize even-parity spin-polarized Fermi surfaces with $d$-, $g$-, or $i$-wave symmetry and are known as altermagnets~\cite{Noda-Nakamura-MomentumdependentBandSpin-2016, Smejkal-Sinova:2020, Hayami-Kusunose-MomentumDependentSpinSplitting-2019, Ahn-Kunes:2019, Yuan-Zunger:2020, Yuan-Zunger-PredictionLowZCollinear-2021, Smejkal-Jungwirth:2022, Smejkal-Jungwirth-ConventionalFerromagnetismAntiferromagnetism-2022, Mazin:2022}. Altermagnets are characterized by broken time-reversal $\mathcal{T}$ symmetry (TRS) but preserved inversion symmetry $\mathcal{P}$; the symmetries originate from the combined lattice geometry and spin ordering in a material leading to momentum-dependent spin splitting~\cite{Noda-Nakamura-MomentumdependentBandSpin-2016, Smejkal-Sinova:2020, Yuan-Zunger:2020, Hayami-Kusunose-BottomupDesignSpinsplit-2020}. The net magnetization of altermagnets vanishes making them distinct from ferromagnets, which also break the TRS.
Many material candidates, such as Ru$_2$O~\footnote{Recent experimental studies suggest that bulk Ru$_2$O may have a nonmagnetic band structure~\cite{Hiraishi-Hiroi-NonmagneticGroundState-2024, Kessler-Moser-Ru02:2024, Wenzel-Tsirlin-FermiliquidBehaviorNonaltermagnetic-2024}; an altermagnetic phase may, however, appear in thin films~\cite{Jeong-Jalan-AltermagneticPolarMetallic-2024}. Therefore, the realization of altermagnetism in Ru$_2$O remains highly controversial.}, MnF$_2$, Mn$_5$Si$_3$, MnTe, and CrSb have been suggested, see also Refs.~\cite{Smejkal-Jungwirth:2022,Bai-Yao-AltermagnetismExploringNew-2024} for a more comprehensive list. The spin-split electron bands were experimentally verified in Refs.~\cite{Fedchenko-Elmers-ObservationTimereversalSymmetry-2024, Krempasky-Jungwirth:2024, Osumi-Sato-ObservationGiantBand-2024, Reimers-Jourdan:2023, Zeng-Liu-ObservationSpinSplitting-2024, Ding-Shen-LargeBandsplittingWave-2024}.

A different class of unconventional magnets can be realized in noncollinear and noncoplanar magnets and features odd-parity, e.g., $p$-wave, structure of magnetization for their itinerant electrons~\cite{Hayami-Kusunose-SpontaneousAntisymmetricSpin-2020, Hayami-Kusunose-BottomupDesignSpinsplit-2020, Hayami-Hayami-MechanismAntisymmetricSpin-2022, Kudasov:2024, Hellenes-Smejkal:2023}. The $p$-wave symmetry of the spin polarization is protected by the $\mathcal{T}\bm{\tau}$ symmetry~\cite{Kudasov:2024, Hellenes-Smejkal:2023}; here, $\bm{\tau}$ denotes a translation, typically by half a unit cell. The $\mathcal{T}\bm{\tau}$ symmetry makes $p$-wave magnets distinct from generic helimagnets, where the $p$-wave spin-polarization can also be observed~\cite{Chen-Ramires:2022, Hayami-Hayami-MechanismAntisymmetricSpin-2022, Kudasov:2024, Mayo-Onose:2024}. As for material candidates, Mn$_3$GaN and CeNiAsO were proposed as platforms for $p$-wave magnetism~\cite{Hellenes-Smejkal:2023}.

Both altermagnets and $p$-wave magnets are characterized by momentum-dependent spin texture and anisotropic Fermi surfaces. Therefore, a natural way to probe their unconventional magnetization is to use local probes or break the inversion symmetry in the system. For example heterostructures with, e.g., planar interfaces provide an ideal setup to study transport properties of altermagnets~\cite{Shao-Tsymbal:2021, Smejkal-Jungwirth-GiantTunnelingMagnetoresistance-2022, Sun-Linder:2023, Papaj:2023, Das-Soori:2023, Ang2023, Samanta-Tsymbal-TunnelingMagnetoresistanceMagnetic-2023, Xu-Zhao-SpinflopMagnetoresistanceCollinear-2023, Giil-Brataas:2024hat, Sukhachov-Linder-ThermoelectricEffectAltermagnetSuperconductor-2024} and $p$-wave magnets~\cite{Maeda-Tanaka:2024, Brekke-Linder-MinimalModelsTransport-2024}. In addition to transport, local probes can also allow one to investigate the unconventional spin-splitting. As we recently demonstrated in Ref.~\cite{Hodt-Linder-InterfaceinducedMagnetizationAltermagnets-2024}, due to the broken inversion symmetry near interfaces, altermagnets develop nonvanishing magnetization near the edges allowing one to glimpse the symmetry and orientation of the spin-polarized Fermi surfaces of an altermagnet in the induced magnetization.

In addition to planar interfaces and edges, point-like defects are both ubiquitous and instrumental in extracting information about the properties of materials. For example, the study of the local density of states (LDOS) in the vicinity of impurities and impurity-induced bound states and resonances in high-$T_c$ superconductors provides information on the structure of their order parameter~\cite{Salkola-Schrieffer:1997, Pan-Davis:2000, Hudson-Davis:2001, Sukhachov-Glazman:2023-impurity} and the shape of the underlying Fermi surface~\cite{Uldemolins-Simon:2022}, see also Refs.~\cite{Balatsky-Zhu:rev-2006, Alloul-Hirschfeld:review-2007} for reviews. The spin-resolved LDOS can be experimentally investigated via scattering tunneling spectroscopy/microscopy (STS/STM)~\cite{Wortmann-Blugel-ResolvingComplexAtomicScale-2001, Wiesendanger:review-2009, Schlenhoff-Wiesendanger-RealspaceImagingAtomicscale-2020}. Because defects break the translation invariance, they allow elastic scattering between the states of the same energy but with different momenta. In real space, the interference of these scattered states is manifested as an intricate modulation of the LDOS. The Fourier transform of such LDOS, known as the quasiparticle interference (QPI) pattern, provides information about the Fermi surface and the allowed scattering channels; the latter are related to spin texture. The corresponding technique is also known as the Fourier-transformed scanning tunneling microscopy (FT-STM)~\cite{Sprunger-Besenbacher-GiantFriedelOscillations-1997, Petersen-Plummer-DirectImagingTwodimensional-1998}. Therefore, to facilitate the application of the well-developed FT-STM technique for unconventional magnets, it is crucial to identify their characteristic features in the impurity effects.

A few of the physical effects caused by impurities in altermagnets have already been studied. In particular, the interaction of two spinful impurities in altermagnets, namely the Ruderman–Kittel–Kasuya–Yosida (RKKY) interaction, was recently investigated in Refs.~\cite{Lee2023e, Amundsen-Linder-RKKYInteractionRashba-2024}. The RKKY interaction becomes anisotropic and shows the characteristic for $d$-wave altermagnets $C_4$-symmetric oscillating pattern. The corresponding pattern is characterized by several periods and is highly directional. In addition, the suppression of the Kondo temperature in altermagnets was predicted in Ref.~\cite{Diniz-Vernek-SuppressedKondoScreening-2024}. However, a comprehensive study of the spatial distribution of the LDOS in altermagnets and, in particular, $p$-wave magnets in the presence of both nonmagnetic and magnetic impurities remains unaddressed.

In this work, we determine the spin-resolved LDOS and magnetization induced by a single impurity with a generic scattering potential that includes both spin-independent and spin-dependent parts. In addition to altermagnets, we consider unconventional $p$-wave magnets focusing on two models available in the literature. Since the band structure of $p$-wave magnets is similar to Rashba metals with an exchange field, we find it instructive to provide a comprehensive study of the LDOS there as well. The results for the Rashba metal are used to emphasize the characteristic features of unconventional magnets.

We find that even a spin-less impurity can induce local magnetization in an altermagnet. The magnetization acquires a characteristic $d$-wave shape reflecting the symmetry of altermagnetic Fermi surfaces. The patterns of charge and spin LDOS have similar four-fold structures but are rotated by $\pi/4$ with respect to each other. The information on the shape of the Fermi surface can be extracted from the periods of the oscillations. Due to their multi-band nature and spin texture, Rashba metals and $p$-wave magnets show a rich structure of charge and spin LDOS that strongly depends on the energy. We demonstrate that an exchange field in a Rashba metal and the $sd$-coupling in one of the models of $p$-wave magnets lift the scattering restrictions imposed by the spin texture and allow the spin splitting to be extracted from the charge LDOS even for a nonmagnetic impurity. This is different from the previous proposals in the context of a Rashba metal that required a magnetic adatom~\cite{Lounis-Blugel-MagneticAdatomInduced-2012}. In these models, we also find that the higher-energy band may allow for the LDOS oscillations with a doubled period compared with the regular Friedel oscillations; the conventional behavior is restored at larger distances from the impurity. We show that this effect originates from the interference of the propagating and evanescent waves corresponding to energy-separated bands and allowed by the spin texture or impurity magnetization. Another feature of the same model of $p$-wave magnets is the direct connection between the anisotropy of the Fermi surface and the spin polarization. This is reflected in different frequencies of the LDOS oscillations along different crystallographic directions. This feature is, however, not universal and is absent in the other model of $p$-wave magnets considered in this work. Our results are instrumental for the study of novel magnets in STS/STM probes and provide an additional way to investigate the band structure of unconventional magnets.

This paper is organized as follows. In Sec.~\ref{sec:model}, we introduce the $T$-matrix formalism and define the key quantities that will be used in the paper. We consider a metal with a Rashba coupling and an exchange field in Sec.~\ref{sec:Rashba}. The LDOS and local magnetization of altermagnets are calculated in Sec.~\ref{sec:AM}. Impurity effects in $p$-wave magnets are analyzed in Sec.~\ref{sec:p-wave}, where we employ two different models. The results are summarized and discussed in Sec.~\ref{sec:Summary}. A few technical aspects related to the Green functions in Rashba metals and $p$-wave magnets are presented in Appendix~\ref{sec:App-1}. Throughout this paper, we use $\hbar=k_{\rm B}=1$.

\section{Formalism and definitions}
\label{sec:model}

We consider a single point-like impurity (e.g., adatom) in a material (substrate). The coupling between the impurity atom and itinerant electrons in the substrate is described by the following Hamiltonian:
\begin{equation}
\label{model-H-int}
H_{\rm imp} = \sum_{\mathbf{k},\mathbf{k}'} \psi_{\mathbf{k},s}^{\dag}\left(V_0 \delta_{s s'} +J \mathbf{S}\cdot \hat{\bm{\sigma}}_{ss'}\right) \psi_{\mathbf{k}',s'}.
\end{equation}
Here, $s=\uparrow, \downarrow$ denotes the spin projection, $\hat{\bm{\sigma}}$ is the electron spin operator, $V_0$ is the strength of potential coupling, $J$ is the strength of exchange coupling, and $\psi_{\mathbf{k} s}$ ($\psi_{\mathbf{k} s}^{\dag}$) annihilates (creates) an electron with the momentum $\mathbf{k}$ and the spin projection $s$. We assume that the impurity spin is classical $\mathbf{S}=S \hat{\mathbf{z}}$. This allows us to introduce the impurity potential as $\hat{V}=V_0 +\sigma_z V_{z}$ with $\sigma_z$ being the Pauli matrix in the spin space and $V_{z} =JS$.

The impurity breaks the translation invariance for itinerant electrons and introduces local modifications of the spectral quantities. We quantify the latter via the spin-resolved LDOS, which is given by
\begin{equation}
\label{model-nu-def}
\nu_{s}(\omega, \mathbf{r}) = -\frac{1}{\pi} \tr{\frac{1+s\sigma_z}{2}\IM{G(\omega,\mathbf{r},\mathbf{r})}},
\end{equation}
where $G(\omega,\mathbf{r},\mathbf{r})$ is the Green's function of the system in the presence of the impurity, $\omega$ is the energy, and $\mathbf{r}$ is the coordinate defined with respect to the impurity position. The charge and spin LDOSs are defined as $\nu_{\rm el}(\omega, \mathbf{r}) = \nu_{\uparrow}(\omega, \mathbf{r}) +\nu_{\downarrow}(\omega, \mathbf{r})$ and $\nu_{\sigma}(\omega, \mathbf{r}) = \nu_{\uparrow}(\omega, \mathbf{r}) -\nu_{\downarrow}(\omega, \mathbf{r})$, respectively.

The local magnetization is obtained by integrating over all occupied states:
\begin{equation}
\label{model-m-def}
\mathbf{m}(\mathbf{r}) = -\frac{g\mu_B}{\pi} \int d\omega\, \tr{\IM{\bm{\sigma} G(\omega,\mathbf{r},\mathbf{r})}} n_{F}(\omega),
\end{equation}
where $g$ is the Land\'{e} factor, $\mu_B$ is Bohr magneton, $\bm{\sigma}$ is the vector of the Pauli matrices in the spin space, and $n_{F}(\omega)$ is the Fermi-Dirac distribution.

To evaluate the Green's function $G(\omega,\mathbf{r},\mathbf{r})$, we employ the T-matrix formalism~\cite{Shiba:1968,Hirschfeld-Einzel:1988}, see also Ref.~\cite{Balatsky-Zhu:rev-2006}. For a point-like impurity, the Green's function reads
\begin{equation}
\label{model-G-R-full}
G(\omega,\mathbf{r},\mathbf{r}) = G_{0}(\omega,\mathbf{r},\mathbf{r}) + G_{0}(\omega,\mathbf{r},\mathbf{0}) T G_{0}(\omega,\mathbf{0},\mathbf{r}),
\end{equation}
where $G_{0}(\omega,\mathbf{r},\mathbf{r}')=G_{0}(\omega,\mathbf{r}-\mathbf{r}',\mathbf{0})$ is the electron Green's function in the absence of the impurity. Its Fourier transform,
\begin{equation}
\label{model-G-k}
G_{0}(\omega,\mathbf{k})= \int d\mathbf{r}\, e^{-i\mathbf{k}\cdot \mathbf{r}} G_{0}(\omega,\mathbf{r},\mathbf{0})
\end{equation}
is defined by the standard relation
\begin{equation}
\label{model-G-0}
G_{0}(\omega,\mathbf{k})= \frac{1}{\omega +i0^{+} -H(\mathbf{k})},
\end{equation}
where $H(\mathbf{k})$ is the Hamiltonian of the clean system.

The $T$-matrix for the local impurity potential is defined as
\begin{equation}
\label{model-T-def}
T = \hat{V} + \hat{V} G_{0}(\omega,\mathbf{0},\mathbf{0}) T.
\end{equation}
Equation (\ref{model-T-def}) is straightforwardly solved:
\begin{equation}
\label{model-T-calc}
T = \left[1 -\hat{V} G_{0}(\omega,\mathbf{0},\mathbf{0}) \right]^{-1} \hat{V}.
\end{equation}

After substituting Eq.~(\ref{model-G-R-full}) into Eq.~(\ref{model-nu-def}), the first term, $G_{0}(\omega,\mathbf{0},\mathbf{0})$, gives rise to the constant DOS of the system (the DOS in the absence of the impurity) and the second term, $G_{0}(\omega,\mathbf{r},\mathbf{0}) T G_{0}(\omega,\mathbf{0},\mathbf{r})$, describes the LDOS modulations $\delta\nu_s(\omega,\vecr)$ due to the impurity. We underline that this solution for $G(\omega,\vecr,\vecr)$ is exact and nonperturbative.

In the limit of the weak scattering potential, we can replace $T\to \hat{V}$. Then, the oscillating part of the LDOS $\delta \nu_s(\omega, \mathbf{r})$ reads
\begin{eqnarray}
\label{model-nu-delta}
&&\lim_{\hat{V}\to0} \delta \nu_{s}(\omega, \mathbf{r}) \nonumber\\
&&= -\frac{1}{\pi} \tr{\frac{1+s\sigma_z}{2} \IM{G_0(\omega,\mathbf{r},\mathbf{0})}\hat{V}\RE{G_0(\omega,\mathbf{0},\mathbf{r})}}\nonumber\\
&&+ \left(\mbox{Im} \leftrightarrow \mbox{Re} \right).
\end{eqnarray}

In the following sections, we apply the described formalism to Rashba metals, altermagnets, and $p$-wave magnets.

\section{Rashba metal}
\label{sec:Rashba}

To provide a background for our studies of impurity effects in unconventional magnets, we find it instructive to calculate the LDOS around an impurity in a Rashba metal with an exchange field. The corresponding results will be contrasted with those in altermagnets and, in particular, $p$-wave magnets, whose model turns out to resemble that of a Rashba metal with spin-splitting, see Sec.~\ref{sec:p-wave-2}.

The interplay of the Rashba spin-orbital coupling (SOC) and impurity effects was investigated in Refs.~\cite{Petersen-Hedegard-SimpleTightbindingModel-2000, Walls-Heller-MultiplescatteringTheoryTwodimensional-2006, Walls-Heller-SpinOrbitCoupling-2007}. It was found that the splitting of the bands due to pure Rashba SOC cannot be captured by the charge LDOS for a single impurity. Quantum corrals composed of several nonmagnetic atoms, on the other hand, lead to more complicated patterns and allow for SOC-induced modulation to be manifested. Impurity-induced spin textures and orbital magnetization in a Rashba electron gas with a magnetic impurity were studied in Refs.~\cite{Lounis-Blugel-MagneticAdatomInduced-2012, Bouaziz-Lounis-ImpurityinducedOrbitalMagnetization-2018}. Being magnetic, even a single adatom is enough to access the Rashba splitting via the spin LDOS.

We use the following model of a 2D Rashba metal~\cite{Rashba:1960,Bychkov-Rashba-OscillatoryEffectsMagnetic-1984}:
\begin{equation}
\label{Rashba-H}
H(\mathbf{k}) = \frac{k^2}{2m} -\mu + \alpha\left(\sigma_x k_y -\sigma_y k_x\right) + h \sigma_z,
\end{equation}
where $k=|\mathbf{k}|$, $m$ is the effective mass, $\mu$ is the chemical potential, $\alpha$ is the strength of the Rashba SOC, and $h$ is the exchange field.

We find it convenient to use dimensionless variables:
\begin{equation}
\label{Rashba-dimless}
\tilde{h} = \frac{h}{\mu}, \quad \quad \tilde{\alpha} = \frac{2m \alpha}{k_F}, \quad \quad \tilde{k} = \frac{k}{k_F}, \quad \tilde{r} = r k_F \,\, \mbox{and} \,\, \hat{\tilde{V}} = \hat{V} \nu_0
\end{equation}
with $k_F=\sqrt{2m\mu}$ being the Fermi momentum in the absence of the SOC and $\nu_0=m/(2\pi)$ being the DOS of the clean system in the absence of the exchange field and the spin splitting. Other dimensionless quantities are introduced in the same way, i.e., $\tilde{\omega}=\omega/\mu$ and $\tilde{\epsilon}=\epsilon/\mu$.

The energy spectrum of the model (\ref{Rashba-H}) is isotropic
\begin{equation}
\label{Rashba-eps}
\tilde{\epsilon}_{\pm} = \tilde{k}^2-1 \pm \sqrt{\tilde{h}^2 +\tilde{\alpha}^2k^2}.
\end{equation}
At a given energy $\tilde{\omega}$, we find the following momenta corresponding to the upper ($\tilde{k}_{+}$) and lower ($\tilde{k}_{-}$) bands:
\begin{equation}
\label{Rashba-k}
\tilde{k}_{\pm}^2 = \tilde{\omega}+1 +\frac{\tilde{\alpha}^2}{2} \mp \sqrt{\tilde{h}^2 +\frac{\tilde{\alpha}^4}{4} +\tilde{\alpha}^2(\tilde{\omega}+1)}.
\end{equation}

\begin{figure}[t!]
\centering
\includegraphics[width=0.9\columnwidth]{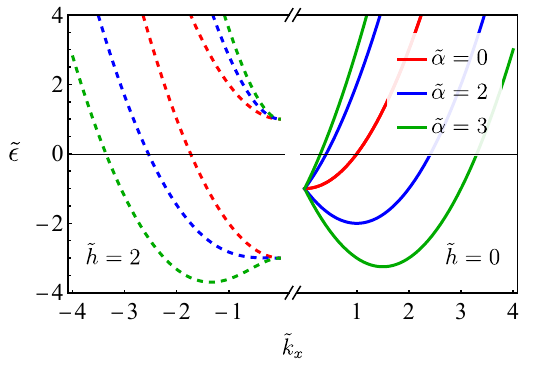}
\caption{\label{fig:Rashba-eps}
The energy spectrum (\ref{Rashba-eps}) at $\tilde{k}_y=0$ where solid and dashed lines correspond to $\tilde{h}=0$ and $\tilde{h}=2$. Note that the value of the exchange field $\tilde{h}=2$ is large for a metal and is used to highlight the features of the model. The spectrum has rotational symmetry and is thus symmetric under $\tilde{\mathbf{k}} \to -\tilde{\mathbf{k}}$.
}
\end{figure}

The energy spectrum (\ref{Rashba-eps}) is shown in Fig.~\ref{fig:Rashba-eps}. The spectrum develops local minima for $|\tilde{\alpha}|^2>2|\tilde{h}|$ leading to the annulus-shaped Fermi surface.

\subsection{Green's functions and analytical results}
\label{sec:Rashba-Green}

The key element of the T-matrix approach described in Sec.~\ref{sec:model} is the Green function of the clean system. The dimensionless retarded Green's function (\ref{model-G-0}) for the model (\ref{Rashba-H}) is
\begin{eqnarray}
\label{Rashba-G-0}
&&G_{0}(\tilde{\omega},\tilde{\mathbf{k}}) \nonumber\\
&&=\frac{\tilde{\omega}+1 -\tilde{k}^2 +\tilde{h} \sigma_z +\tilde{\alpha} \left(\sigma_x \tilde{k}_y -\sigma_y \tilde{k}_x\right)}{\left(\tilde{\omega}+1-\tilde{k}^2\right)^2 -\tilde{h}^2 -\tilde{\alpha}^2 \tilde{k}^2 +i0^{+}\sign{\tilde{\omega}+1 -\tilde{k}^2}}. \nonumber\\
\end{eqnarray}

In calculating the inverse Fourier transform, we separate the principal value part and the imaginary part via the Sokhotski–Plemelj formula,
\begin{eqnarray}
\label{Rashba-G-0-r}
G_{0}(\tilde{\omega},\tilde{\mathbf{r}},\mathbf{0}) &=&
-2\nu_0 \int d\tilde{k} \tilde{k} \int_0^{2\pi} \frac{d\theta}{2\pi} e^{i\tilde{k} \tilde{r} \cos{\theta}} \nonumber\\
&\times& \Bigg\{\mbox{p.v.} \frac{\tilde{\omega}+1 -\tilde{k}^2 +\tilde{h} \sigma_z +\tilde{\alpha} \left(\sigma_x \tilde{k}_y -\sigma_y \tilde{k}_x\right)}{(\tilde{k}^2-\tilde{k}_{+}^2)(\tilde{k}^2-\tilde{k}_{-}^2)} \nonumber\\
&+&i \pi \sign{\tilde{\omega}+1 -\tilde{k}^2} \df{(\tilde{k}^2 -\tilde{k}_{+}^2)(\tilde{k}^2 -\tilde{k}_{-}^2)} \nonumber\\
&\times& \left[\tilde{\omega}+1 -\tilde{k}^2 +\tilde{h} \sigma_z +\tilde{\alpha} \left(\sigma_x \tilde{k}_y -\sigma_y \tilde{k}_x\right)\right]\Bigg\}, \nonumber\\
\end{eqnarray}
where p.v. denotes the principal value. The explicit form of the real-space Green's functions is given in Eqs.~(\ref{Rashba-G-r0}) and (\ref{Rashba-G-r}).

The LDOS and the magnetization are calculated via Eqs.~(\ref{model-nu-def}) and (\ref{model-m-def}) with the full Green's function given in Eq.~(\ref{model-G-R-full}). Since the LDOS for a scattering potential of an arbitrary magnitude is cumbersome to study analytically, we focus on the case of a weak scattering potential, $T\approx \hat{V}$, and extract the key qualitative aspects of the impurity-induced oscillations. Numerical results are presented in Sec.~\ref{sec:Rashba-numerics}.

According to Eq.~(\ref{model-nu-delta}), the oscillating part of the LDOS is determined by the product of the real and imaginary part of the Green's functions. By using Eq.~(\ref{Rashba-G-r}), we arrive at the following expression for the oscillating part of the charge LDOS:
\begin{widetext}
\begin{equation}
\label{Rashba-nu-delta-0}
\delta \nu_{\rm el}(\tilde{\omega},\tilde{\mathbf{r}}) \propto \!\! \sum_{\rho_1,\rho_2=\pm} \rho_1 \sign{\tilde{\omega}+1 -\tilde{k}_{\rho_2}^2} \left\{\left[\left(\tilde{\omega}+1-\tilde{k}_{\rho_1}^2\right)\left(\tilde{\omega}+1-\tilde{k}_{\rho_2}^2\right) +\tilde{h}^2\right] Y_0{\left(\tilde{k}_{\rho_1}\tilde{r}\right)}J_0{\left(\tilde{k}_{\rho_2}\tilde{r}\right)} +\tilde{\alpha}^2 \tilde{k}_{\rho_1}\tilde{k}_{\rho_2} Y_1{\left(\tilde{k}_{\rho_1}\tilde{r}\right)}J_1{\left(\tilde{k}_{\rho_2}\tilde{r}\right)}\right\},
\end{equation}
\end{widetext}
where $J_n(x)$ and $Y_n(x)$ are the Bessel functions of the first and second kinds, respectively.
The same equation albeit with the sign minus at the second term is obtained for the spin LDOS $\delta \nu_{\sigma}(\tilde{\omega},\tilde{\mathbf{r}})$ around a magnetic impurity with $\hat{V}=V_z \sigma_z$.
In writing Eq.~(\ref{Rashba-nu-delta-0}), we omitted an overall prefactor that is determined by the difference of the momenta and does not depend on the band indices $\rho_{1,2}$.

Assuming the limit $\tilde{r}\to \infty$, setting $\tilde{h}=0$, and using Eqs.~(\ref{Rashba-k}), (\ref{Rashba-nu-delta-0}), (\ref{App-Rashba-J-rinf}), and (\ref{App-Rashba-Y-rinf}), one can show that $\delta \nu_{\rm el}(\tilde{\omega},\tilde{\mathbf{r}}) \propto \tilde{V}_0\cos{\left((\tilde{k}_{+} +\tilde{k}_{-})\tilde{r}\right)}/\tilde{r}$. Therefore, the charge LDOS oscillates with a single frequency and decays as $1/\tilde{r}$ away from the impurity. This result agrees with that in Refs.~\cite{Petersen-Hedegard-SimpleTightbindingModel-2000, Walls-Heller-MultiplescatteringTheoryTwodimensional-2006, Lounis-Blugel-MagneticAdatomInduced-2012}. As discussed in Ref.~\cite{Petersen-Hedegard-SimpleTightbindingModel-2000}, the single frequency of the oscillations originates from the spin texture of the helical bands where the scattering between the states $\tilde{k}_{\pm}$ and $-\tilde{k}_{\pm}$ is forbidden. This results in the absence of the oscillations determined by $2\tilde{k}_{+}$ and $2\tilde{k}_{-}$. The effect of the nontrivial spin texture is taken automatically in the Green's function approach.

In the case of a magnetic impurity and spin LDOS, $\delta \nu_{\sigma}(\tilde{\omega},\tilde{\mathbf{r}}) \propto \tilde{V}_z \tilde{\alpha}^2 \left[\cos{\left(2\tilde{k}_{+}\tilde{r} \right)} +\cos{\left(2\tilde{k}_{-}\tilde{r} \right)}\right]/\tilde{r}$. Hence, there are oscillations at two frequencies corresponding to scatterings between the same helical bands, enabled by the spin-flip processes now permitted by the magnetic impurity.

The exchange field $\tilde{h}$ changes the structure and the spin texture of the helical bands, see Fig.~\ref{fig:Rashba-eps}, lifting the scattering restriction. Then, as follows from Eq.~(\ref{Rashba-nu-delta-0}), there is no cancellation between oscillating terms resulting in the charge and spin LDOS oscillations at the frequencies determined by $2\tilde{k}_{\pm}$ and $\tilde{k}_{+}+\tilde{k}_{-}$:
\begin{widetext}
\begin{eqnarray}
\label{Rashba-nu-delta-el}
\delta \nu_{\rm el}(\tilde{\omega},\tilde{\mathbf{r}}) &\propto&
\tilde{V}_0\frac{\tilde{h}^2}{2} \frac{\cos{(2\tilde{k}_{+}\tilde{r})} +\cos{(2\tilde{k}_{-}\tilde{r})}}{\tilde{r}} -\tilde{V}_0\left[\tilde{\alpha}^2 \sqrt{(\tilde{\omega}+1)^2 -\tilde{h}^2} +\tilde{\alpha}^2(\tilde{\omega}+1) +\tilde{h}^2\right] \frac{\cos{\left((\tilde{k}_{+}+\tilde{k}_{-})\tilde{r}\right)}}{\tilde{r}},\\
\label{Rashba-nu-delta-sigma}
\delta \nu_{\sigma}(\tilde{\omega},\tilde{\mathbf{r}}) &\propto&
\tilde{V}_z \sum_{\rho=\pm}\frac{2\tilde{\alpha}^2 \tilde{k}^2_{\rho} +\tilde{h}^2}{2} \frac{\cos{(2\tilde{k}_{\rho}\tilde{r})}}{\tilde{r}}
+\tilde{V}_z \left[\tilde{\alpha}^2 \sqrt{(\tilde{\omega}+1)^2 -\tilde{h}^2} -\tilde{\alpha}^2(\tilde{\omega}+1) -\tilde{h}^2\right] \frac{\cos{\left((\tilde{k}_{+}+\tilde{k}_{-})\tilde{r}\right)}}{\tilde{r}}.
\end{eqnarray}
\end{widetext}

Another nontrivial statement that can be extracted from Eq.~(\ref{Rashba-nu-delta-0}), is the presence of the LDOS oscillations with the doubled period when the bottom of the higher band is approached (we assume $\tilde{h}\neq0$). These oscillations are observed in the vicinity of the impurity, $|\tilde{k}_{+}| \tilde{r} \lesssim 1$, and vanish at larger distances, i.e., at $|\tilde{k}_{+}| \tilde{r} \gg 1$. Such oscillations originate from the interference of the propagating (i.e., with $\tilde{k}_{-}^2>0$) and evanescent (i.e., with $\tilde{k}_{+}^2<0$) waves.

\begin{figure*}[ht!]
\centering
\begin{subfloat}[]{\includegraphics[height=0.64\columnwidth]{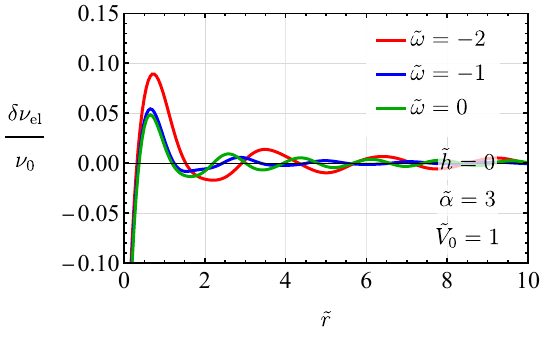}}
\end{subfloat}
\begin{subfloat}[]{\includegraphics[height=0.64\columnwidth]{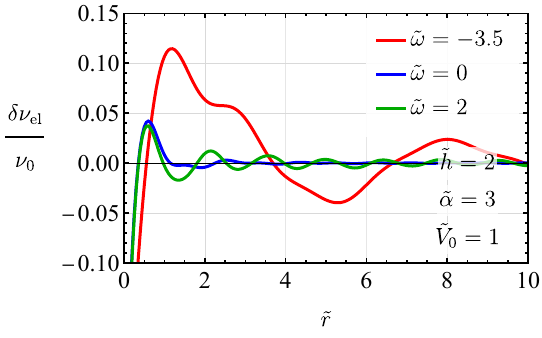}}
\end{subfloat}\\
\begin{subfloat}[]{\includegraphics[height=0.64\columnwidth]{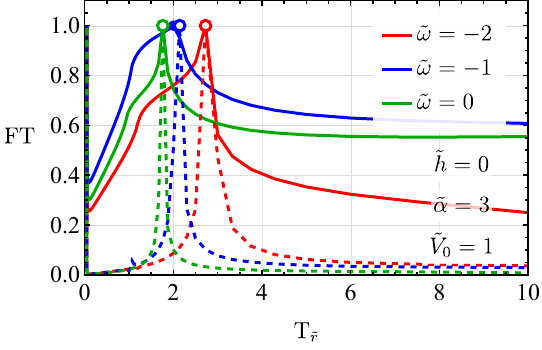}}
\end{subfloat}
\begin{subfloat}[]{\includegraphics[height=0.64\columnwidth]{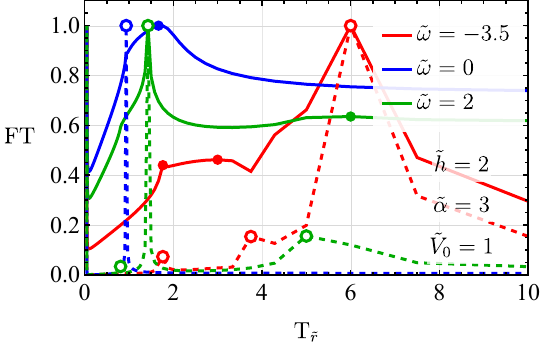}}
\end{subfloat}
\caption{\label{fig:Rashba-LDOS-charge}
The oscillating part of the charge LDOS (panels (a) and (b)) as a function of $\tilde{r}$ and its Fourier transform (panels (c) and (d)). We fix a few characteristic energies, see Fig.~\ref{fig:Rashba-eps} for the shape of the energy dispersion relation. We use $\tilde{h}=0$ in panels (a) and (c) and $\tilde{h}=2$ in panels (b) and (d). Solid and dashed lines in panels (c) and (d) show the relative contribution to the real-space oscillations of the LDOS from harmonics with different periods $T_{\tilde{r}}$ near the impurity ($0<\tilde{r}<30$) and away from it ($50<\tilde{r}<80$), respectively, obtained via a Fourier transformation (FT). Filled and empty circles mark the corresponding maxima. In all panels, we assume a nonmagnetic impurity with $\hat{V} = V_0$ and $\tilde{V}_0=1$ as well as set $\tilde{\alpha}=3$.
The exact expression for the $T$-matrix is used in numerical calculations.
}
\end{figure*}

Indeed, near the bottom of the upper band, $|\tilde{k}_{+}| \ll 1$ and $\tilde{k}_{+}^2<0$, hence
\begin{equation}
Y_n(|\tilde{k}_{+}| \tilde{r}) \to 2K_n(|\tilde{k}_{+}| \tilde{r})/\pi.
\end{equation}
Assuming $|\tilde{k}_{-}| \tilde{r} \gg 1$ and $|\tilde{k}_{+}| \tilde{r} \lesssim 1$ in Eq.~(\ref{Rashba-nu-delta-0}), it is straightforward to show that there will be terms $\propto \sin{\left(\tilde{k}_{-} \tilde{r} -\pi/4\right)}$ corresponding to the quasi-harmonic oscillations with doubled period. These terms are exponentially suppressed at $|\tilde{k}_{+}|\tilde{r}>1$ where $K_n(|\tilde{k}_{+}| \tilde{r}) \sim e^{-|\tilde{k}_{+}| \tilde{r}}$.

The nontrivial spectrum of the Rashba model with two energy bands and local minima for $|\tilde{\alpha}|^2>2|\tilde{h}|$ allows to realize three interesting regimes based solely on the shape of the constant-energy contours: (i) annulus-shaped constant-energy contours at $|\tilde{\alpha}|^2>2|\tilde{h}|$ and $\tilde{\omega}<-1-\tilde{h}$; (ii) single circle-like constant-energy contour at $-1-\tilde{h}<\tilde{\omega}<-1+\tilde{h}$ and any value of $\tilde{\alpha}$; and (iii) two circle-like constant-energy contours at $\tilde{\omega}>-1+\tilde{h}$.

In the case of the annulus-shaped constant-energy contours, $\tilde{k}_{\pm}^2>0$, hence there should be oscillations with the periods determined by $2\tilde{k}_{\pm}$ and $|\tilde{k}_{+}+\tilde{k}_{-}|$. As we discuss above, at $\tilde{h}=0$, only the charge oscillations determined by $|\tilde{k}_{+}+\tilde{k}_{-}|$ remain. The spin LDOS, on the other hand, is determined by $2\tilde{k}_{\pm}$.
The same qualitative behavior is observed for the case with two constant-energy contours, $\tilde{\omega}>-1+\tilde{h}$.

The case with a single constant-energy contour at $-1-\tilde{h}<\tilde{\omega}<-1+\tilde{h}$ is characterized by the LDOS oscillations dictated by $2\tilde{k}_{-}$. However, when the energies approach the bottom of the higher band, the LDOS in the vicinity of the impurity $|\tilde{k}_{+}|\tilde{r}\lesssim 1$ also shows oscillations at doubled period; see also the discussion after Eq.~(\ref{Rashba-nu-delta-sigma}). It is interesting that the higher band can still influence the LDOS oscillations via evanescent waves despite being unoccupied.

As we will show in Sec.~\ref{sec:p-wave-2}, the doubling of the oscillations period near the impurity due to interference of propagating and evanescent waves is a generic phenomenon that goes beyond the particular model of a Rashba metal. Technically, the phenomenon originates from the mixing of the states corresponding to different bands in $\tr{G_0(\omega,\mathbf{r},\mathbf{0})\hat{V} G_0(\omega,\mathbf{0},\mathbf{r})}$. The mixing can be of intrinsic, i.e., inherent to a particular model Hamiltonian, or extrinsic, i.e., induced by the impurity, nature. In the intrinsic case, the necessary ingredients are the separation of the bands in energy and their nontrivial spin (or pseudospin) structure.  In terms of the Green functions, the mixing of the states can be related to the diagonalization of the Green functions. If there are no spin-dependent interactions and the spin texture is trivial, Green's functions are diagonalized via a momentum-independent unitary transformation, which leaves $\tr{G_0(\omega,\mathbf{r},\mathbf{0})\hat{V} G_0(\omega,\mathbf{0}, \mathbf{r})}$ in the oscillating part of the LDOS invariant. Therefore, only the product of diagonal terms enters the final result for the LDOS signifying the absence of the cross-band interference. On the other hand, if the spin texture is nontrivial, as is the case for the Rashba model (\ref{Rashba-H}), the diagonalization requires a momentum-dependent transformation. In this case, we cannot eliminate the contribution of the cross-terms in $\tr{G_0(\omega,\mathbf{r}, \mathbf{0})\hat{V} G_0(\omega,\mathbf{0}, \mathbf{r})}$, which results in an interference between waves corresponding to different bands crucial for the period doubling. We underline that this can happen even when one of these waves is evanescent, meaning that one of the bands is unoccupied. The requirement of the nontrivial spin polarization can be relaxed for a magnetic impurity leading to an extrinsic mechanism of the state mixing. For example, this is the case for the model (\ref{Rashba-H}) at $\tilde{\alpha}=0$ and $\tilde{h}\neq0$ with the impurity potential $\hat{V}=V_y\sigma_y$. In our work, we focus on the intrinsic mechanism.

\subsection{Numerical results}
\label{sec:Rashba-numerics}

To support our analytical considerations, we present numerical results for the LDOS oscillations in the Rashba metal both with and without the exchange field $\tilde{h}$. We use Eqs.~(\ref{model-nu-def}) and (\ref{model-G-R-full}) where the real-space Green functions are given in Eqs.~(\ref{Rashba-G-r0}) and (\ref{Rashba-G-r}). The oscillating part of the charge LDOS is shown in Fig.~\ref{fig:Rashba-LDOS-charge} for a few energies that encompass different regions of the dispersion relation, see Fig.~\ref{fig:Rashba-eps}. To identify the dominant harmonics, we perform the Fourier transform of the oscillating part of the LDOS $\delta \nu(\tilde{\omega}, \tilde{r})$ and extract the corresponding period of oscillations $T_{\tilde{r}}$.

\begin{figure*}[ht!]
\centering
\begin{subfloat}[]{\includegraphics[height=0.64\columnwidth]{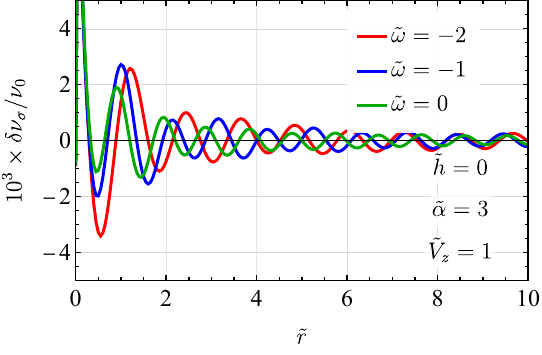}}
\end{subfloat}
\begin{subfloat}[]{\includegraphics[height=0.64\columnwidth]{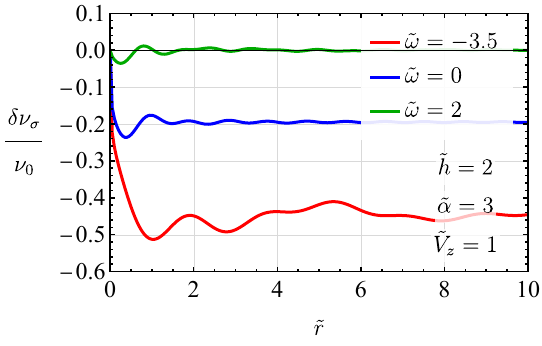}}
\end{subfloat}\\
\begin{subfloat}[]{\includegraphics[height=0.6\columnwidth]{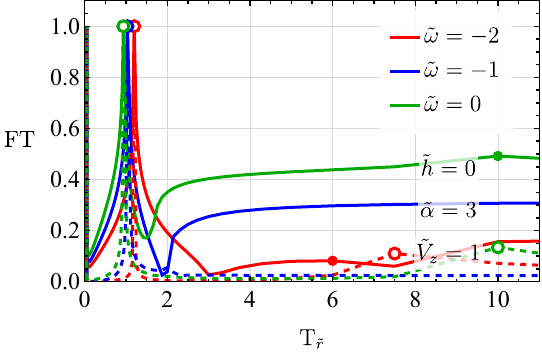}}
\end{subfloat}
\begin{subfloat}[]{\includegraphics[height=0.64\columnwidth]{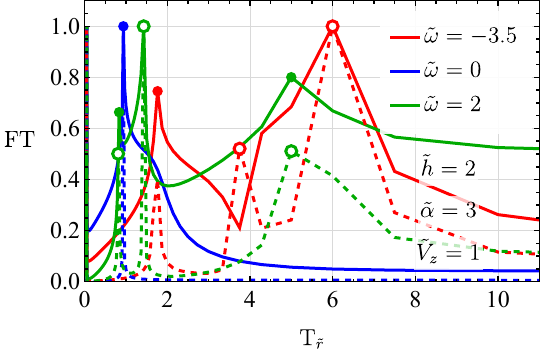}}
\end{subfloat}
\caption{\label{fig:Rashba-LDOS-spin}
The oscillating part of the spin LDOS (panels (a) and (b)) as a function of $\tilde{r}$ and its Fourier transform (panels (c) and (d)). We fix a few characteristic energies, see Fig.~\ref{fig:Rashba-eps} for the shape of the energy dispersion relation. We use $\tilde{h}=0$ in panels (a) and (c) and $\tilde{h}=2$ in panels (b) and (d). Solid and dashed lines in panels (c) and (d) show the relative contribution to the real-space oscillations of the spin LDOS from harmonics with different periods $T_{\tilde{r}}$ near the impurity ($0<\tilde{r}<30$) and away from it ($50<\tilde{r}<80$), respectively, obtained via the FT. In all panels, we assume a magnetic impurity with $\hat{V} = \sigma_z V_z$ and $\tilde{V}_z=1$ as well as set $\tilde{\alpha}=3$.
The exact expression for the $T$-matrix is used in numerical calculations.
}
\end{figure*}

Without the exchange field, there is only one dominant frequency of oscillations determined by the sum of characteristic momenta $|\tilde{k}_{+}+\tilde{k}_{-}|$ at given energy, see Figs.~\ref{fig:Rashba-LDOS-charge}(a) and \ref{fig:Rashba-LDOS-charge}(c). The spin LDOS for a magnetic impurity shows the oscillations with two dominant frequencies determined by $2\tilde{k}_{+}$ and $2\tilde{k}_{-}$, see Figs.~\ref{fig:Rashba-LDOS-spin}(a) and \ref{fig:Rashba-LDOS-spin}(c). These findings confirm the qualitative discussion about the role of the spin texture in the previous section, see also Ref.~\cite{Petersen-Hedegard-SimpleTightbindingModel-2000}.

Adding the exchange field to the system makes the oscillation pattern more complicated with additional well-pronounced peaks in the Fourier transform, see Figs.~\ref{fig:Rashba-LDOS-charge}(b) and \ref{fig:Rashba-LDOS-charge}(d) for the charge LDOS around a nonmagnetic impurity and Figs.~\ref{fig:Rashba-LDOS-spin}(b) and \ref{fig:Rashba-LDOS-spin}(d) for the spin LDOS around a magnetic impurity. In agreement with our theoretical discussion after Eq.~(\ref{Rashba-nu-delta-0}), the peaks corresponds to $2\tilde{k}_{+}$, $2\tilde{k}_{-}$, and $|\tilde{k}_{+}+\tilde{k}_{-}|$.

\begin{figure}[ht!]
\centering
\begin{subfloat}[]{\includegraphics[width=0.99\columnwidth]{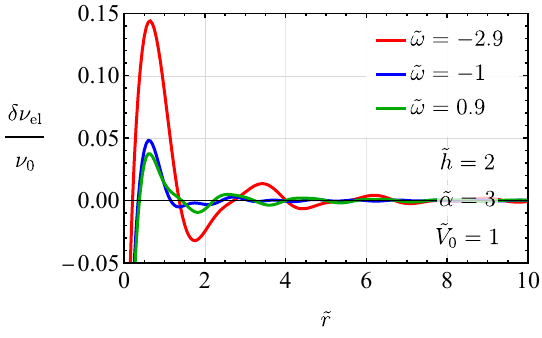}}
\end{subfloat}
\begin{subfloat}[]{\includegraphics[width=0.99\columnwidth]{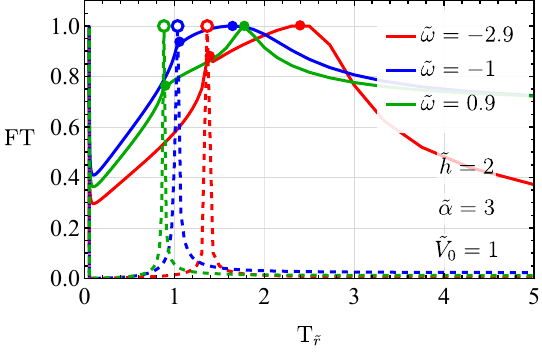}}
\end{subfloat}
\caption{\label{fig:Rashba-LDOS-two-band}
The oscillating part of the charge  LDOS (panel (a)) as a function of $\tilde{r}$ and its Fourier transform (panel (b)). We fix a few characteristic energies near the bottom of the lower band (red lines) and approaching the bottom of the higher band (blue and green lines). Solid and dashed lines in panel (b) show the relative contribution to the real-space oscillations of the charge LDOS from the harmonics with different periods $T_{\tilde{r}}$ at $0<\tilde{r}<30$ and $50<\tilde{r}<80$, respectively. Filled and empty circles mark the corresponding maxima. In all panels, we use the spin-independent impurity potential with $\tilde{V}_0=1$ as well as set $\tilde{h}=2$ and $\tilde{\alpha}=3$.
The exact expression for the $T$-matrix is used in numerical calculations.
}
\end{figure}

Finally, let us demonstrate another interesting property of the Rashba model at a nonzero exchange field, namely the appearance of the LDOS oscillations in the vicinity of impurity with a doubled period. In agreement with our discussion after Eq.~(\ref{Rashba-nu-delta-0}), the charge LDOS oscillations shown in Fig.~\ref{fig:Rashba-LDOS-two-band} do not have a single well-defined period. The peaks in the Fourier transform correspond to $2\tilde{k}_{-}$ and $\tilde{k}_{-}$. The latter (i.e., defined by $\tilde{k}_{-}$) disappears if only the long-range tail of the LDOS is considered, cf. solid and dashed lines in Fig.~\ref{fig:Rashba-LDOS-two-band}.

Summarizing this section, we find that subjecting a Rashba metal to an external spin-splitting field provides an alternative way to probe the Rashba spin splitting in the LDOS oscillations even around a nonmagnetic impurity. In addition, we uncover unusual LDOS oscillations at a doubled period allowed by the two-band structure of the model. Later, we will see that this phenomenon also occurs for $p$-wave magnets even in the absence of an external spin-splitting field.

\section{Altermagnets}
\label{sec:AM}

In this section, we discuss the impurity-induced LDOS oscillations in an altermagnet. We use the following dimensionless Hamiltonian:
\begin{equation}
\label{AM-H}
H(\tilde{\mathbf{k}}) = \tilde{k}^2 -1 +\sigma_z \tilde{k}^2 J(\theta),
\end{equation}
where $J(\theta)$ defines the altermagnetic spin splitting. We assume a $d$-wave symmetry of the splitting and, without the loss of generality, choose the coordinate system such that $J(\theta) = t_{\rm AM} \cos{(2\theta)}$ with $t_{\rm AM}<1$.

\begin{figure}[ht!]
\centering
\includegraphics[width=0.99\columnwidth]{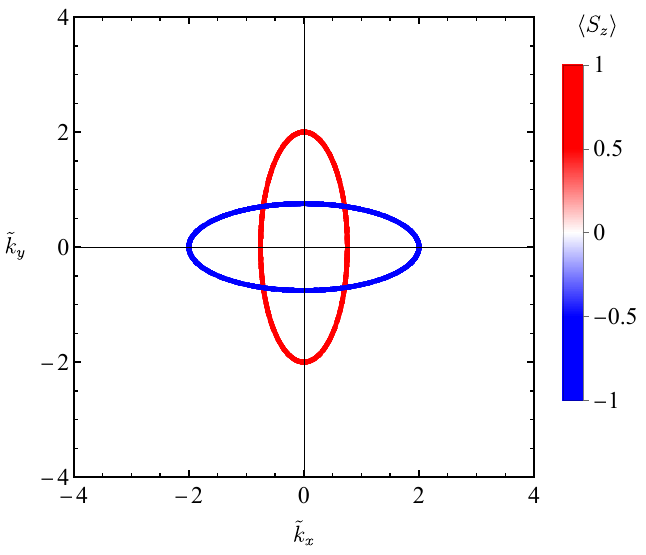}
\caption{
\label{fig:AM-FS}
Constant energy contour for a $d$-wave altermagnet at $\tilde{\omega} = 0$ and $t_{\rm AM} =0.75$. We use Eq.~(\ref{AM-eps}). Red and blue colors correspond to the spin projection $\left\langle S_z\right\rangle$ of the fully spin-polarized bands in an altermagnet.
}
\end{figure}

The energy spectrum of the Hamiltonian (\ref{AM-H}) is
\begin{equation}
\label{AM-eps}
\tilde{\epsilon}_s = \tilde{k}^2 \left[1+sJ(\theta)\right] -1,
\end{equation}
where $s=\pm$ is the spin projection, and is visualized in Fig.~\ref{fig:AM-FS}.

The structure of the model (\ref{AM-H}), i.e., the full spin polarization, significantly simplifies the calculation of the Green's function $G_{0}(\tilde{\omega},\tilde{\mathbf{k}}) = \mbox{diag}{\left\{G_{+, 0}(\tilde{\omega},\tilde{\mathbf{k}}), G_{-, 0}(\tilde{\omega},\tilde{\mathbf{k}})\right\}}$, whose components read
\begin{equation}
\label{AM-a-G-0}
G_{s, 0}(\tilde{\omega},\tilde{\mathbf{k}})= \frac{1}{\tilde{\Omega} +i0^{+}  -\tilde{k}^2\left[1 +s J(\theta)\right]^2},
\end{equation}
where $\tilde{\Omega} = \tilde{\omega} +1$. By introducing the new variable $\mathbf{q}_s = \left\{\tilde{k}_x \sqrt{1+st_{\rm AM}}, \tilde{k}_y \sqrt{1-st_{\rm AM}}\right\}$, we bring the altermagnetic Green's function to a form resembling the  Green's function in a regular metal. The real-space Green's functions are
\begin{eqnarray}
\label{AM-a-G-r0-0}
G_{s, 0}(\tilde{\omega},\mathbf{0},\mathbf{0}) &=&  -\frac{\nu_0}{1-t_{\rm AM}^2} \ln{\left(\frac{\tilde{\Lambda}^2}{|\tilde{\Omega}|}\right)} -i\frac{\pi \nu_0}{1-t_{\rm AM}^2} \Theta{(\tilde{\Omega})},\\
\label{AM-a-G-r-0}
G_{s, 0}(\tilde{\omega},\tilde{\mathbf{r}},\mathbf{0}) &=& \frac{\pi \nu_0 \Theta{(\tilde{\Omega})}}{1-t_{\rm AM}^2} \left[Y_{0}{\left(\sqrt{\tilde{\Omega}}\, \tilde{r}_s(\theta_r)\right)} -iJ_{0}{\left(\sqrt{\tilde{\Omega}}\, \tilde{r}_s(\theta_r)\right)}\right] \nonumber\\
&-& \frac{2\nu_0 \Theta{(-\tilde{\Omega})}}{1-t_{\rm AM}^2} K_{0}{\left(\sqrt{|\tilde{\Omega}|}\, \tilde{r}_s(\theta_r)\right)},
\end{eqnarray}
where we introduced a shorthand
\begin{equation}
\label{AM-a-rho}
\tilde{r}_s(\theta_r) = \tilde{r} \sqrt{\frac{\cos^2{\theta_r}}{1+s t_{\rm AM}} +\frac{\sin^2{\theta_r}}{1-st_{\rm AM}}}
\end{equation}
and $\theta_r$ is the angle between $\tilde{\mathbf{r}}$ and $\hat{\mathbf{x}}$.

Since $\tilde{r}_s(\theta_r)$ is spin-dependent, we expect a nontrivial spin LDOS even for a nonmagnetic impurity. Indeed, assuming weak scattering potential, $\tilde{V}_s=|V_s|\nu_0 \ll1$, we obtain the following oscillating part, see also Eq.~(\ref{model-nu-delta}), of the spin-resolved LDOS:
\begin{eqnarray}
\label{AM-a-nu-app}
\delta\nu_s(\tilde{\omega}, \tilde{\mathbf{r}}) \!\! &=& \!\! \frac{2\pi \tilde{V}_s \nu_0}{1-t_{\rm AM}^2} Y_{0}{\left(\sqrt{\tilde{\Omega}}\, \tilde{r}_s \right)} J_{0}{\left(\sqrt{\tilde{\Omega}}\, \tilde{r}_s\right)} \nonumber\\
\!\!&\stackrel{\sqrt{\tilde{\Omega}} \tilde{r} \gg1}{\approx}&\!\!
-\frac{2\tilde{V}_s \nu_0}{(1-t_{\rm AM}^2) \sqrt{\tilde{\Omega}}\, \tilde{r}_s(\theta_r)} \cos{\left(2\sqrt{\tilde{\Omega}}\, \tilde{r}_s(\theta_r)\right)}, \nonumber\\
\end{eqnarray}
which allows for $\sum_{s=\pm} \delta\nu_s(\tilde{\omega}, \tilde{\mathbf{r}}) \neq0$ even at $V_s=V_0$. Only for $\theta_r=\pi/4+\pi n$ with $n\in \mathds{Z}$, i.e., at the ``altermagnetic nodes", the spin-resolved LDOS around a nonmagnetic impurity vanishes. To show this explicitly, we expand in small $t_{\rm AM}$ up to the first order and keep the leading order terms in $1/(\sqrt{\tilde{\Omega}} \tilde{r})$:
\begin{equation}
\label{AM-a-nu-app-tAM}
\delta\nu_s(\tilde{\omega}, \tilde{\mathbf{r}}) \approx -\frac{2\tilde{V}_s \nu_0}{\sqrt{\tilde{\Omega}}\,\tilde{r}} \left[1 + \frac{s t_{\rm AM}}{2} \cos{(2\theta_r)}\right]\cos{\left(2\sqrt{\tilde{\Omega}}\, \tilde{r}\right)}.
\end{equation}

As follows from Eqs.~(\ref{AM-a-rho}), (\ref{AM-a-nu-app}), and (\ref{AM-a-nu-app-tAM}), the periods of both charge and spin LDOS have characteristic fourfold directional dependence whose magnitude depends on the altermagnetic splitting $t_{\rm AM}$. Furthermore, for all directions except $\theta_r=\pi/4+\pi n$, there are two periods corresponding to two spin-polarized constant energy contours, see Fig.~\ref{fig:AM-FS}.

\begin{figure}[ht!]
\centering
\begin{subfloat}[]{\includegraphics[width=0.99\columnwidth]{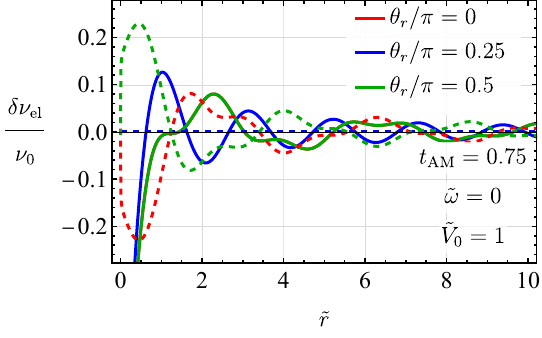}}
\end{subfloat}
\begin{subfloat}[]{\includegraphics[width=0.99\columnwidth]{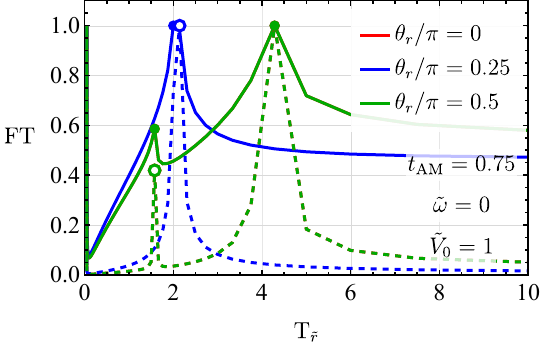}}
\end{subfloat}
\caption{\label{fig:AM-LDOS-theta}
The oscillating part of the charge (solid lines) and spin (dashed lines) LDOS as a function of $\tilde{r}$ are shown in panel (a). The Fourier transform of the charge LDOS is shown in panel (b). Solid and dashed lines in panel (b) show the relative contribution to the real-space oscillations of the charge LDOS from the harmonics with different periods $T_{\tilde{r}}$ in the coordinate ranges $0<\tilde{r}<30$ (solid lines) and $50<\tilde{r}<80$ (dashed lines). We assumed a nonmagnetic impurity with the potential $\tilde{V}_0=1$ as well as fixed $t_{\rm AM}=0.75$ and $\tilde{\omega}=0$.
The exact expression for the $T$-matrix is used in numerical calculations.
}
\end{figure}

We visualize the oscillating part of the charge and spin LDOS and determine the corresponding frequencies in Fig.~\ref{fig:AM-LDOS-theta}. The charge LDOS has a four-fold symmetry and is the same at $\theta_r=0$ and $\theta_r=\pi/2$; note that red ($\theta_r=0$) and green ($\theta_r=\pi/2$) lines in Fig.~\ref{fig:AM-LDOS-theta} overlap. The spin LDOS at $\theta_r=0$ and $\theta_r=\pi/2$ have the same oscillation periods but opposite signs. The Fourier transform of the oscillations clearly shows two well-pronounced peaks in any direction except the direction along the altermagnetic nodes where there is only one peak.

\begin{figure}[ht!]
\centering
\begin{subfloat}[]{\includegraphics[width=0.99\columnwidth]{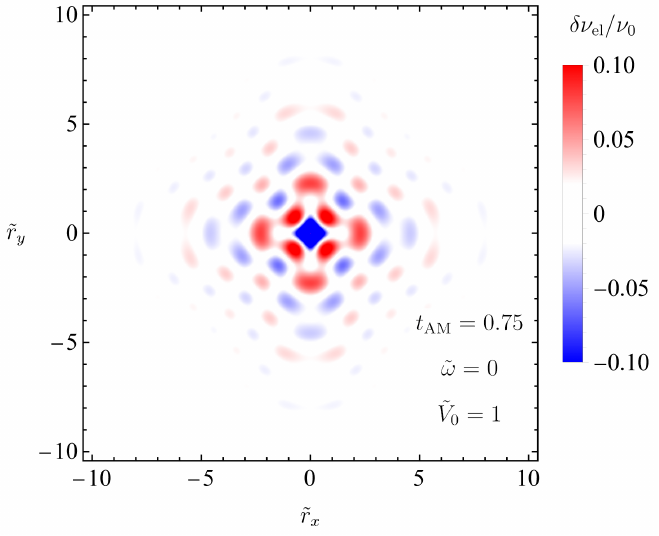}}
\end{subfloat}
\begin{subfloat}[]{\includegraphics[width=0.99\columnwidth]{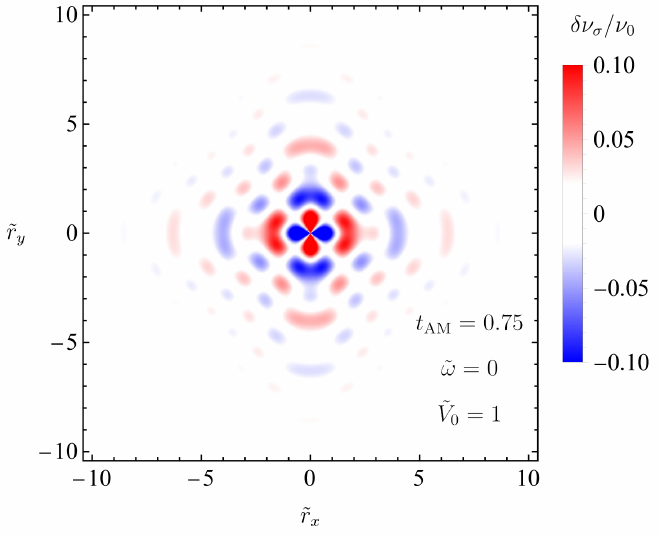}}
\end{subfloat}
\caption{\label{fig:AM-LDOS-density}
The oscillating part of the charge (panel (a)) and spin (panel (b)) LDOS as the function of the radial coordinates $\tilde{r}_x$ and $\tilde{r}_y$. The LDOS is the same for the local $\tilde{V}_s=\tilde{V}_0=1$ and spinfull $\tilde{V}_s=s\tilde{V}_z=s$ impurity potential. In all panels, we set $\tilde{\omega}=0$ and $t_{\rm AM}=0.75$.
The exact expression for the $T$-matrix is used in numerical calculations.
}
\end{figure}

The LDOS as a function of $\tilde{\mathbf{r}}$ is shown in Fig.~\ref{fig:AM-LDOS-density}. While the periods of LDOS oscillations are the same for the charge and spin LDOS, the patterns are different. Unlike the spin LDOS, whose maxima coincide with the principal axes of the altermagnet, the maxima of the charge LDOS are rotated by $\pi/4$ and are directed along the nodal lines of the altermagnet, i.e., the lines at which there is no spin splitting, cf. Figs.~\ref{fig:AM-FS} and \ref{fig:AM-LDOS-density}. Still, both densities show characteristic four-fold symmetry. A similar structure of the LDOS was also predicted and observed in $d$-wave superconductors, see Ref.~\cite{Balatsky-Zhu:rev-2006} for a review.

\begin{figure}[ht!]
\centering
\begin{subfloat}[]{\includegraphics[width=0.99\columnwidth]{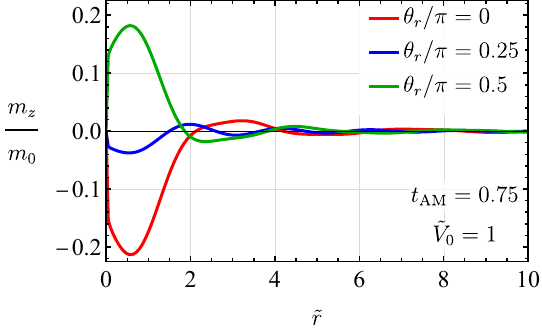}}
\end{subfloat}
\begin{subfloat}[]{\includegraphics[width=0.99\columnwidth]{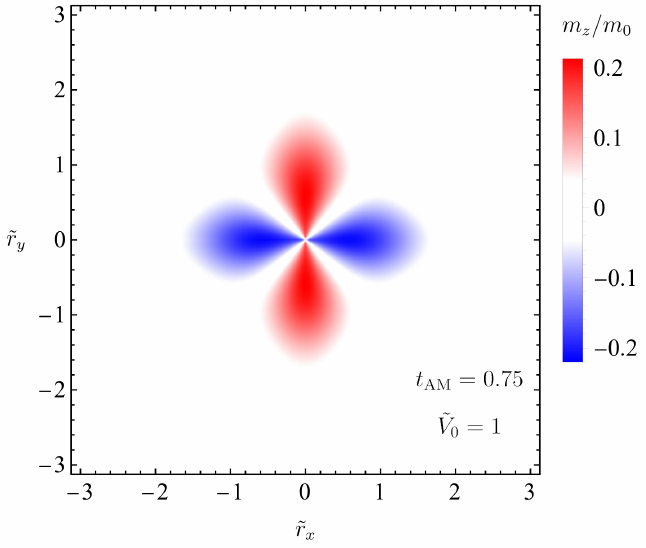}}
\end{subfloat}
\caption{\label{fig:AM-m}
Local magnetization $m_z(\tilde{\mathbf{r}})$ defined in Eq.~(\ref{model-m-def}) as a function of the distance from the impurity $\tilde{\mathbf{r}}$ in a $d$-wave altermagnet.
In all panels, we use $m_0 = g\mu_B \nu_0 \mu$, $t_{\rm AM}=0.75$, and $\tilde{V}_s=\tilde{V}_0=1$.
The exact expression for the $T$-matrix is used in numerical calculations.
}
\end{figure}

As we demonstrated above, magnetic and nonmagnetic impurities can induce oscillating spin LDOS in altermagnets. For a strong nonmagnetic impurity $\tilde{V}_0\to\infty$, the corresponding induced magnetization is determined exclusively by the altermagnetic spin-splitting. Therefore, the measurements of the local magnetization could provide another glimpse into altermagnetic properties. We present the local magnetization $m_z(\tilde{\mathbf{r}})$ defined by Eq.~(\ref{model-m-def}) in Fig.~\ref{fig:AM-m}. As expected, the oscillating magnetization inherits the $d$-wave symmetry and spin polarization of the Fermi surfaces of the altermagnet and shows faster decay with $\tilde{r}$ compared to the LDOS, i.e., $m_z(\tilde{\mathbf{r}}) \propto 1/\tilde{r}^2$. Indeed, integrating the expression (\ref{AM-a-nu-app}) over all occupied states, we obtain
\begin{equation}
\label{AM-a-m-app}
\delta m_z(\tilde{\mathbf{r}}) \approx -2\sum_{s=\pm}\frac{s\tilde{V}_s \nu_0}{(1-t_{\rm AM}^2) \tilde{r}_s^2(\theta_r)} \sin{\left(2\tilde{r}_s(\theta_r)\right)}.
\end{equation}
Assuming $t_{\rm AM}\ll1$ and $\tilde{V}_s= \tilde{V}_0$, we derive
\begin{equation}
\label{AM-a-m-app-1}
\delta m_z(\tilde{\mathbf{r}}) \approx -\frac{4\tilde{V}_0 \nu_0}{\tilde{r}^2} \sin{\left(2\tilde{r}\right)} \cos{\left(2\theta_r\right)},
\end{equation}
which explicitly demonstrates the $d$-wave pattern of the magnetization.

To summarize this section, a pattern of the LDOS around both nonmagnetic and magnetic defects in altermagnets provides a direct way to map the characteristic altermagnetic spin splitting. By measuring the LDOS pattern and the local magnetization, both the symmetry and the size of the constant energy contours can be extracted. In comparison with the Rashba metal with an exchange field, altermagnets lack energy splitting between the bands, which precludes a regime where one of the fully spin-polarized bands is occupied and the other is empty. This leads to a simpler structure of the LDOS patterns with no more than two dominating periods.

\section{\texorpdfstring{$p$-wave}{p-wave} magnets}
\label{sec:p-wave}

In this section, we investigate the LDOS in unconventional $p$-wave magnets.

\subsection{Model I}
\label{sec:p-wave-1}

We start with a simpler model of $p$-wave magnets proposed in Ref.~\cite{Hayami-Kusunose-BottomupDesignSpinsplit-2020} and used in Ref.~\cite{Maeda-Tanaka:2024}. The model features two fully spin-polarized bands and is described by the following Hamiltonian:
\begin{equation}
\label{p-wave-1-H}
H(\tilde{\mathbf{k}}) = \tilde{k}^2-1  + \sigma_z \left(\tilde{\bm{\alpha}}\cdot\tilde{\mathbf{k}}\right).
\end{equation}

\begin{figure}[ht!]
\centering
\includegraphics[width=0.99\columnwidth]{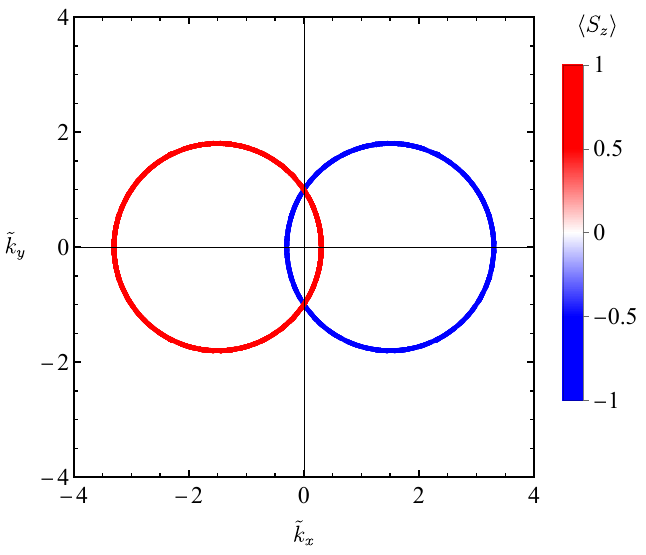}
\caption{
\label{fig:p-wave-1-FS-energy}
Constant energy contour at $\tilde{\omega} = 0$ and $\tilde{\alpha}=3$. We use the $p$-wave Hamiltonian~(\ref{p-wave-1-H}). Red and blue colors correspond to the spin projection $\left\langle S_z\right\rangle$ of the fully spin-polarized bands in a $p$-wave magnet.
}
\end{figure}

The energy spectrum is visualized in Fig.~\ref{fig:p-wave-1-FS-energy} and represents two decoupled fully spin-polarized parabolic bands.

As in altermagnets, the full spin polarization of the model (\ref{p-wave-1-H}) allows to represent the Green's function (\ref{model-G-0}) in a simple form $G_{0}(\tilde{\omega},\tilde{\mathbf{k}}) = \mbox{diag}{\left\{G_{+, 0}(\tilde{\omega},\tilde{\mathbf{k}}), G_{-, 0}(\tilde{\omega},\tilde{\mathbf{k}})\right\}}$ with
\begin{equation}
\label{p-wave-1-G-0}
G_{s, 0}(\tilde{\omega},\tilde{\mathbf{k}}) =  \frac{1}{\tilde{\Omega} +i0^{+} -q_s^2},
\end{equation}
where $\mathbf{q}_s = \tilde{\mathbf{k}} +s \tilde{\bm{\alpha}}/2$ and  $\tilde{\Omega} = \tilde{\omega} +1 +\tilde{\alpha}^2/4$.

The inverse Fourier transform of Green's functions are similar to those in an altermagnet, see Eqs.~(\ref{AM-a-G-r0-0}) and (\ref{AM-a-G-r-0}). In particular, $G_{s, 0}(\tilde{\omega},\mathbf{0},\mathbf{0})$ is given by Eq.~(\ref{AM-a-G-r0-0}) at $t_{\rm AM}=0$. As for $G_{s, 0}(\tilde{\omega},\tilde{\mathbf{r}},\mathbf{0})$, one needs to replace $t_{\rm AM}\to0$ and $\tilde{r}_s(\theta_r) \to \tilde{r}$, as well as multiply by $e^{is (\tilde{\bm{\alpha}}\cdot \tilde{\mathbf{r}})}$. The latter phase factor is, however, irrelevant in the LDOS; see the last term in Eq.~(\ref{model-G-R-full}) where the phases $e^{is (\tilde{\bm{\alpha}}\cdot \tilde{\mathbf{r}})}$ cancel.

\begin{figure*}[ht!]
\centering
\begin{subfloat}[]{\includegraphics[height=0.4\columnwidth]{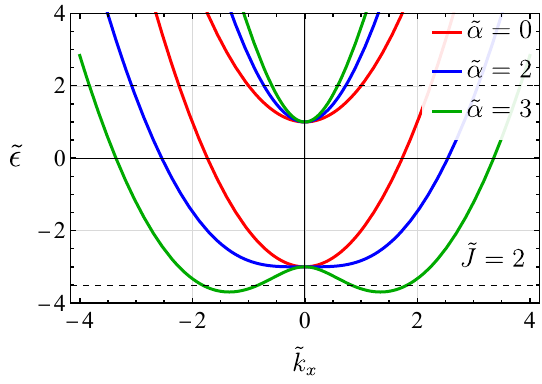}}
\end{subfloat}
\begin{subfloat}[$\tilde{\epsilon}=-3.1$]{\includegraphics[height=0.4\columnwidth]{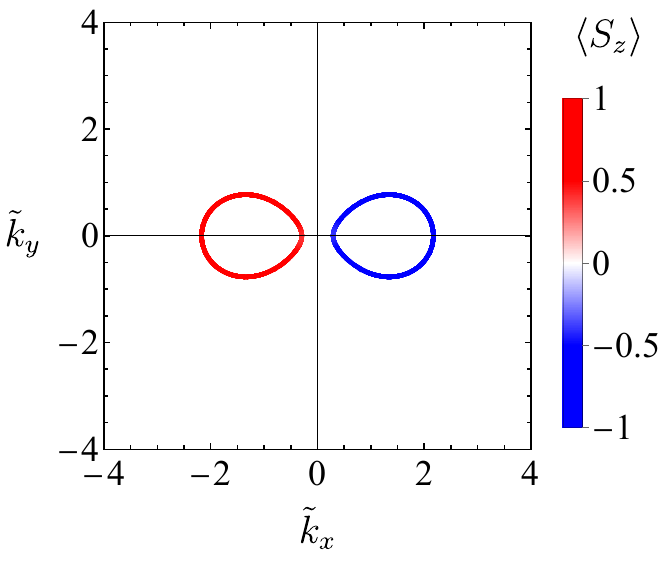}}
\end{subfloat}
\begin{subfloat}[$\tilde{\epsilon}=0$]{\includegraphics[height=0.4\columnwidth]{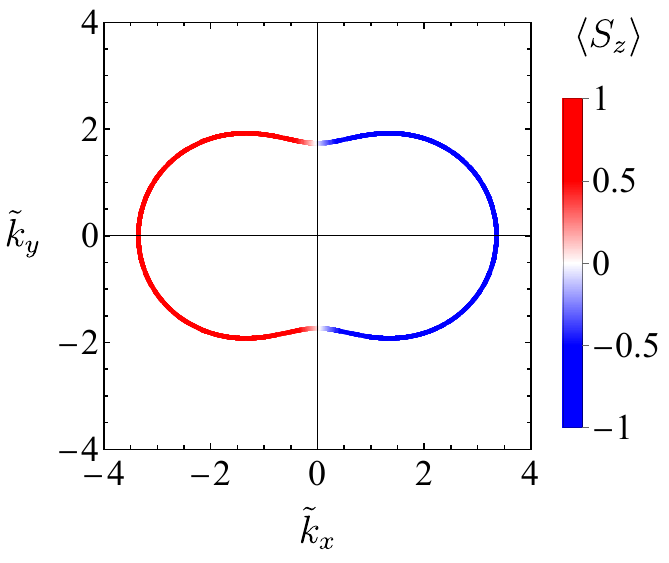}}
\end{subfloat}
\begin{subfloat}[$\tilde{\epsilon}=2$]{\includegraphics[height=0.4\columnwidth]{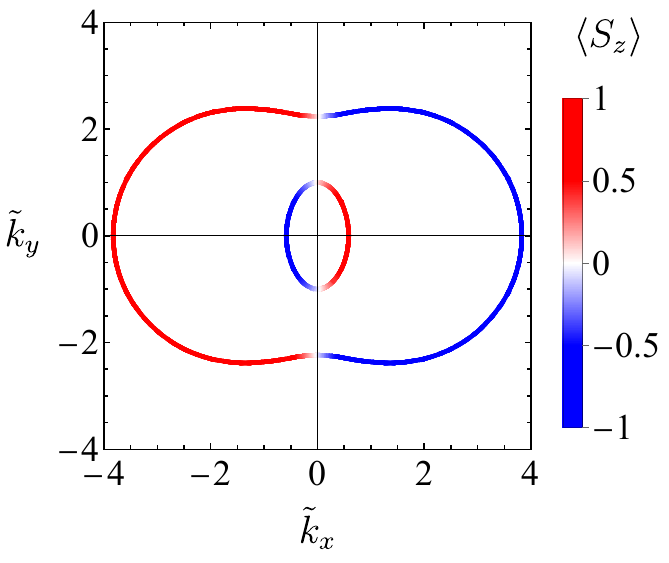}}
\end{subfloat}
\caption{\label{fig:p-wave-2-eps}
Panel (a): The energy spectrum of the $p$-wave magnet model defined in Eq.~(\ref{p-wave-2-eps}) at $\tilde{k}_y=0$ at a few values of $\tilde{\alpha}$. Panels (b)--(c): Constant energy contours at a few different energies which are shown by horizontal dashed lines in panel (a). We used $\tilde{\alpha}=3$. Red and blue colors correspond to the spin projection $\left\langle S_z\right\rangle$. In all panels, we fix $\tilde{J}=2$ and $\tilde{\bm{\alpha}} \parallel \hat{\mathbf{x}}$.
}
\end{figure*}

\begin{figure*}[ht!]
\centering
\begin{subfloat}[]{\includegraphics[height=0.60\columnwidth]{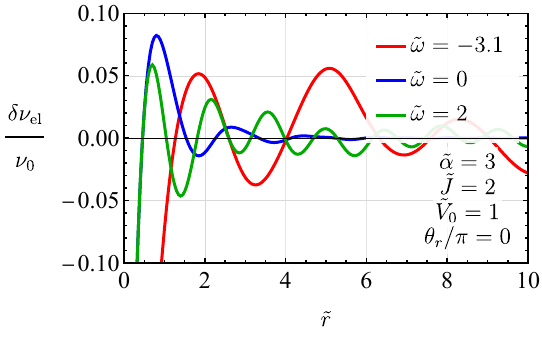}}
\end{subfloat}
\begin{subfloat}[]{\includegraphics[height=0.60\columnwidth]{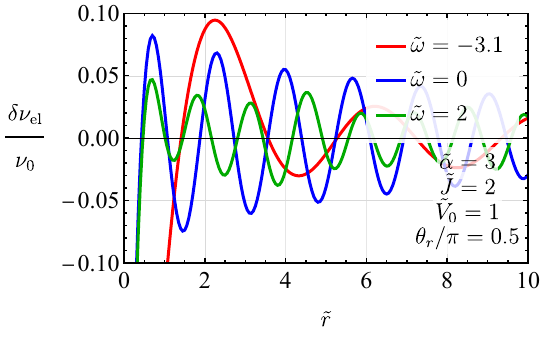}}
\end{subfloat}\\
\begin{subfloat}[]{\includegraphics[height=0.64\columnwidth]{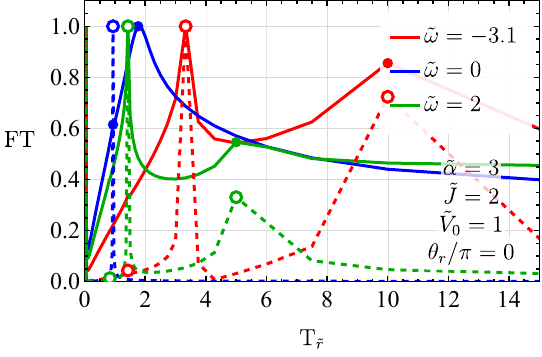}}
\end{subfloat}
\begin{subfloat}[]{\includegraphics[height=0.64\columnwidth]{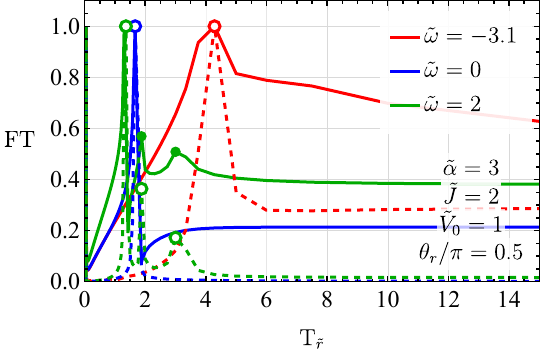}}
\end{subfloat}
\caption{\label{fig:p-wave-2-LDOS-all}
The oscillating parts of the charge LDOS (panels (a) and (b)) in the $p$-wave magnet model (\ref{p-wave-2-H}). The Fourier transform of the charge oscillations (panels (c) and (d)) with $T_{\tilde{r}}$ being the period of each of the contributing harmonics. Solid and dashed lines correspond to the relative contributions to the real-space oscillations of LDOS from the harmonics with different periods $T_{\tilde{r}}$ in the coordinate ranges $0<\tilde{r}<30$ (solid lines) and $30<\tilde{r}<60$ (dashed lines). Panels (a) and (c) show the results at $\tilde{\mathbf{r}}\parallel\tilde{\bm{\alpha}}$ ($\theta_r=0$), and panels (b) and (d) correspond to $\tilde{\mathbf{r}}\perp\tilde{\bm{\alpha}}$ ($\theta_r=\pi/2$). The charge and spin LDOS are calculated for $\hat{\tilde{V}}=\tilde{V}_0=1$ and $\hat{\tilde{V}}=\sigma_z\tilde{V}_z=\sigma_z$, respectively. In all panels, we set $\tilde{\alpha}=3$ and $\tilde{J}=2$.
The exact expression for the $T$-matrix is used in numerical calculations.
}
\end{figure*}

\begin{figure*}[ht!]
\centering
\begin{subfloat}[]{\includegraphics[height=0.64\columnwidth]{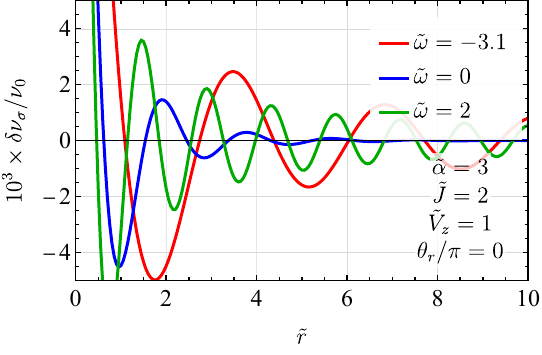}}
\end{subfloat}
\begin{subfloat}[]{\includegraphics[height=0.64\columnwidth]{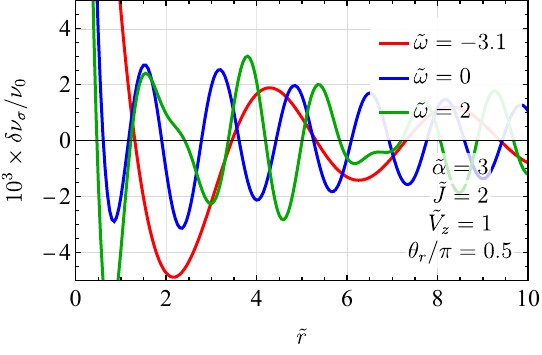}}
\end{subfloat}\\
\begin{subfloat}[]{\includegraphics[height=0.64\columnwidth]{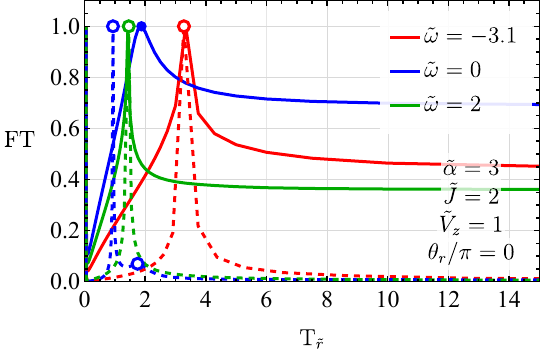}}
\end{subfloat}
\begin{subfloat}[]{\includegraphics[height=0.64\columnwidth]{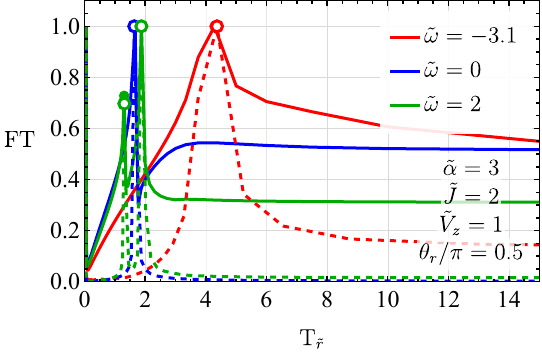}}
\end{subfloat}
\caption{\label{fig:p-wave-2-LDOS-all-spin}
The oscillating parts of the spin LDOS (panels (a) and (b)) in the $p$-wave magnet model (\ref{p-wave-2-H}). The Fourier transform of the spin oscillations (panels (c) and (d)) with $T_{\tilde{r}}$ being the period of each of the contributing harmonics. Solid and dashed lines correspond to the relative contributions to the real-space oscillations of LDOS from the harmonics with different periods $T_{\tilde{r}}$ in the coordinate ranges $0<\tilde{r}<30$ (solid lines) and $30<\tilde{r}<60$ (dashed lines). Panels (a) and (c) show the results at $\tilde{\mathbf{r}}\parallel\tilde{\bm{\alpha}}$ ($\theta_r=0$), and panels (c) and (d) correspond to $\tilde{\mathbf{r}}\perp\tilde{\bm{\alpha}}$ ($\theta_r=\pi/2$). The charge and spin LDOS are calculated for $\hat{\tilde{V}}=\tilde{V}_0=1$ and $\hat{\tilde{V}}=\sigma_z\tilde{V}_z=\sigma_z$, respectively. In all panels, we set $\tilde{\alpha}=3$ and $\tilde{J}=2$.
The exact expression for the $T$-matrix is used in numerical calculations.
}
\end{figure*}

Assuming weak scattering potential, $|\tilde{V}_s| \ll1$, we obtain the following oscillating part of the spin-resolved LDOS:
\begin{eqnarray}
\label{p-wave-1-nu-app}
\delta\nu_s(\tilde{\omega}, \tilde{\mathbf{r}}) &\approx& 2\pi \tilde{V}_s \nu_0 Y_{0}{\left(\sqrt{\tilde{\Omega}}\, \tilde{r}\right)} J_{0}{\left(\sqrt{\tilde{\Omega}}\, \tilde{r}\right)} \nonumber\\
&\stackrel{\sqrt{\tilde{\Omega}} \tilde{r} \gg1}{\approx}& -2\frac{\tilde{V}_s \nu_0}{\sqrt{\tilde{\Omega}}\, \tilde{r}} \cos{\left(2\sqrt{\tilde{\Omega}}\, \tilde{r}\right)},
\end{eqnarray}
cf. Eq.~(\ref{AM-a-nu-app}). Therefore the oscillations of the LDOS decay as $1/\tilde{r}$ away from the impurity, and are described by the single frequency $2\sqrt{\tilde{\Omega}}$. As expected from the preserved in $p$-wave magnets TRS, the spin LDOS requires a magnetic impurity.

Another interesting feature of the model (\ref{p-wave-1-H}), is the isotropy of the corresponding LDOS, i.e., there is no dependence on the direction between $\tilde{\bm{\alpha}}$ and $\tilde{\mathbf{r}}$. This is explained by the absence of the inter-band processes in the model at hand. Therefore, the relative position of the spin-up and spin-down bands is irrelevant.
On the other hand, the spin-splitting parameter $\tilde{\alpha}$ still affects the frequency of oscillating LDOS and magnetization with the latter being $\propto \sin{\left(2\tilde{r} \sqrt{1+\tilde{\alpha}^2/4}\right)}/\tilde{r}^2$; see also Eqs.~(\ref{AM-a-m-app}) and (\ref{AM-a-m-app-1}) at $t_{\rm AM} \to 0$ and $\tilde{r}_s(\theta_r)\to \tilde{r}$.

Therefore, an impurity-induced LDOS does not allow one to directly probe the spin-splitting in the model (\ref{p-wave-1-H}). This conclusion resembles that for a Rashba metal, albeit holds even for a magnetic impurity.

\subsection{Model II}
\label{sec:p-wave-2}

Guided by symmetry arguments, we proposed~\cite{Brekke-Linder-MinimalModelsTransport-2024} a different model of a $p$-wave magnet that is consistent with the $\mathcal{T}\bm{\tau}$ symmetry. Our model is defined by the following Hamiltonian
\begin{equation}
\label{p-wave-2-H}
H_{\eta}(\tilde{\mathbf{k}}) = \tilde{k}^2-1 + \left(\tilde{\bm{\alpha}}\cdot\tilde{\mathbf{k}}\right)\sigma_z +\eta \tilde{J} \sigma_x,
\end{equation}
where $\tilde{\bm{\alpha}}$ defines the spin splitting and $\tilde{J}$ is the dimensionless $sd$ coupling. This model has a richer structure and features two doubly degenerate bands. The degenerate bands are distinguished by the sectoral index $\eta=\pm$. Physically, the sectoral index originates from coarse-graining in a lattice model with a helical magnetization texture describing $p$-wave magnetism: neighboring lattice sites with spins rotated by $\pi/2$ are gathered into pairs dubbed sectors~\cite{Brekke-Linder-MinimalModelsTransport-2024}. The sectors are consistent with the $\mathcal{T}\bm{\tau}$ symmetry: the TRS $\mathcal{T}$ flips the spin, and the half-unit cell translation $\bm{\tau}$ in the lattice model changes the sectoral index $\eta \to -\eta$ thus describing the transformation from one sector to the other.

The energy spectrum of the model (\ref{p-wave-2-H}) is
\begin{equation}
\label{p-wave-2-eps}
\tilde{\epsilon}_{\pm} =  \tilde{k}^2-1 \pm \sqrt{\tilde{J}^2 +\left(\tilde{\bm{\alpha}}\cdot \tilde{\mathbf{k}}\right)^2}.
\end{equation}
Without the loss of generality, we set $\tilde{\bm{\alpha}} \parallel \hat{\mathbf{x}}$. Comparing the energy spectra of the Rashba model (\ref{Rashba-eps}) and the $p$-wave magnet (\ref{p-wave-2-eps}), we note that the key difference is in the angular dependence of the spin-splitting term $\left(\tilde{\bm{\alpha}}\cdot \tilde{\mathbf{k}}\right)$.

The bands feature nontrivial spin polarization (in the units of $\hbar/2$), which is defined as
\begin{equation}
\label{p-wave-2-Sz}
\left<S_z\right> = \pm \frac{2\left(\tilde{\bm{\alpha}}\cdot \tilde{\mathbf{k}}\right)}{\sqrt{\tilde{J}^2 + \left(\tilde{\bm{\alpha}}\cdot \tilde{\mathbf{k}}\right)^2}},
\end{equation}
where the sign $\pm$ corresponds to the sign in Eq.~(\ref{p-wave-2-eps}). For each of the sectors, there is also nonzero $\left<S_x\right>$, which, however, has opposite signs in different sectors. Therefore, after taking into account both sectors, only $\left<S_z\right>$ remains. Note that while the polarization $\langle S_z\rangle$ is odd in momentum, $\langle S_x\rangle$ is even.

We present the energy spectrum and the constant energy contours in Fig.~\ref{fig:p-wave-2-eps}. Unlike the Rashba model used in Sec.~\ref{sec:Rashba}, the band dispersion is anisotropic. The spin texture is also different, i.e., the spin polarization is no longer complete.

The momentum-space Green's function for the Hamiltonian (\ref{p-wave-2-H}) is
\begin{equation}
\label{p-wave-2-G-0}
G_{\eta, 0}(\tilde{\omega},\tilde{\mathbf{k}}) = \frac{\tilde{\omega} +1 -\tilde{k}^2 +\left(\tilde{\bm{\alpha}}\cdot \tilde{\mathbf{k}}\right) \sigma_z +\eta \tilde{J} \sigma_x}{(\tilde{k}^2 -i0^{+} -\tilde{k}_{+}^2)(\tilde{k}^2 -i0^{+}-\tilde{k}_{-}^2)},
\end{equation}
where
\begin{widetext}
\begin{equation}
\label{p-wave-2-a-k-pm}
\tilde{k}_{\pm}^2 = \tilde{\omega} +1 +\frac{\tilde{\alpha}^2 \cos^2{\theta}}{2} \mp \sqrt{\tilde{J}^2 +\frac{\tilde{\alpha}^4\cos^4{\theta}}{4} +\tilde{\alpha}^2(1+\tilde{\omega}) \cos^2{\theta}},
\end{equation}
\end{widetext}
cf. with Eqs.~(\ref{Rashba-G-0}) and (\ref{Rashba-k}).

Because of the anisotropy, the inverse Fourier transform of Green's function (\ref{p-wave-2-G-0}) is more cumbersome and the result is less informative. Therefore, we present it in Appendix~\ref{sec:App-1-p-wave}, and, in what follows, focus on the numerical results.

The oscillating part of the charge and spin LDOS together with their Fourier transforms are shown in Figs.~\ref{fig:p-wave-2-LDOS-all} and \ref{fig:p-wave-2-LDOS-all-spin}. In contrast to the Rashba model used in Sec.~\ref{sec:Rashba}, the oscillations of the LDOS in the model (\ref{p-wave-2-H}) are inherently anisotropic with the periods of oscillation determined by angle-dependent $\tilde{k}_{\pm}$. In particular, the period of the oscillations along $\tilde{\bm{\alpha}}$ (i.e., at $\theta_r=0$) are smaller than that in the direction perpendicular to $\tilde{\bm{\alpha}}$ (i.e., at $\theta_r=\pi/2$); cf. left and right columns in Figs.~\ref{fig:p-wave-2-LDOS-all} and \ref{fig:p-wave-2-LDOS-all-spin}. As in the Rashba model with an exchange field, however, there is a subtle interplay between the propagating and evanescent waves leading to the doubling of the period. This can be seen from the Fourier transform in Figs.~\ref{fig:p-wave-2-LDOS-all}(c) and \ref{fig:p-wave-2-LDOS-all-spin}(c) at $|\tilde{\omega}|\lesssim |\tilde{J}-1|$: there are two peaks for a single constant energy contour with the peak corresponding to a doubled period being dominant. These double-period peaks vanish at $\tilde{r}\to \infty$ (see the dashed lines) confirming their connection to the evanescent waves.

Thus, the key difference between the impurity-induced LDOS oscillations in the $p$-wave model (\ref{p-wave-2-H}) compared to both the Rashba model (\ref{Rashba-H}) and the other $p$-wave model (\ref{p-wave-1-H}) is in the anisotropy of these oscillations. The anisotropy is directly related to the spin splitting, hence we expect different periods of oscillations in magnetic and nonmagnetic phases.  Other properties, including oscillations with three periods as well as the period doubling, are similar to those for the Rashba model with an exchange field; the TRS in the $p$-wave magnet, however, remains preserved.

\section{Summary}
\label{sec:Summary}

In this work, we investigated the LDOS and magnetization around an impurity in materials with a different spin texture including Rashba metals, altermagnets, and $p$-wave magnets. We found that a few characteristic properties of these materials related to the band dispersion and the spin texture can be extracted from the patterns of Friedel oscillation.

In altermagnets, both magnetic and nonmagnetic impurities allow for the oscillating LDOS and local magnetization. The symmetries of the corresponding patterns reflect the symmetry of the altermagnetic splitting, see Eqs.~(\ref{AM-a-nu-app-tAM}) and (\ref{AM-a-m-app-1}) as well as Figs.~\ref{fig:AM-LDOS-density} and \ref{fig:AM-m}. The shape of the constant energy contours in an altermagnet can be inferred from the periods of the LDOS oscillations, see Eq.~(\ref{AM-a-nu-app}) and Fig.~\ref{fig:AM-LDOS-theta}. Therefore the study of the impurity-induced LDOS and magnetization patterns provides an alternative way to quantify the altermagnetic spin splitting.

Our results for $p$-wave magnets revealed a rich structure of the LDOS patterns for a model that is consistent with the defining TRS-translation $\mathcal{T}\bm{\tau}$ symmetry, see Sec.~\ref{sec:p-wave-2}. Unlike the TRS-symmetric Rashba model considered in Sec.~\ref{sec:Rashba}, the spin texture of the $p$-wave model does not provide strong restrictions on the frequencies of the oscillations allowing one to extract the $p$-wave spin splitting even by using a nonmagnetic impurity. In the case of the Rashba metal, the Rashba splitting can be inferred from the LDOS in the presence of a magnetic impurity~\cite{Lounis-Blugel-MagneticAdatomInduced-2012} or, as we demonstrated in this work, by adding an exchange field, see Eqs.~(\ref{Rashba-nu-delta-el}) and (\ref{Rashba-nu-delta-sigma}). An interesting feature of both $p$-wave magnets and magnetized Rashba metals is the LDOS oscillations with a doubled period that originate from the interplay of propagating and evanescent waves corresponding to the split in energy bands. These oscillations are well-pronounced in the vicinity of the impurity when the bottom of the higher-energy band is approached. The relation to the evanescent waves is evident from the decreasing amplitude farther away from the impurity.

The structure of the LDOS patterns around an impurity in a different class of $p$-wave magnets considered in Sec.~\ref{sec:p-wave-1} is much simpler. We found that the impurity-induced LDOS features only a single period and does not allow one to directly probe the spin splitting, see Eq.~(\ref{p-wave-1-nu-app}).

The results of our work are instrumental for the FT-STM and investigation of the QPIs in altermagnets and unconventional magnets. In particular, the LDOS can be probed via spin-polarized STM/STS~\cite{Wortmann-Blugel-ResolvingComplexAtomicScale-2001, Wiesendanger:review-2009, Schlenhoff-Wiesendanger-RealspaceImagingAtomicscale-2020}. Because of the surface sensitivity of these techniques, 2D altermagnets and $p$-wave magnets are the most promising candidates. We mention CrO~\cite{Chen-Sanyal:2021}, FeSe~\cite{Mazin-Smejkal-InducedMonolayerAltermagnetism-2023}, and V$_2$Te$_2$O~\cite{Cui-Yang-GiantSpinHallTunneling-2023}, see also Ref.~\cite{Bai-Yao-AltermagnetismExploringNew-2024} for the list of other 2D altermagnetic candidates.

While in this work we focused on 2D magnetic materials, the qualitative results are also applicable to surfaces of 3D magnets. Another possible extension of this study would be to consider the effect of several impurities, i.e., impurity corrals.

Recently, another independent study addressing impurity scattering and Friedel oscillations in altermagnets appeared in Physical Review B~\cite{Chen-Lou-ImpurityScatteringFriedel-2024}. The results in the latter paper for the anisotropy of the LDOS and different periods of oscillations for spin-up and spin-down electrons agree with those obtained in our work. The cases of $p$-wave magnets and Rashba metals, however, were not analyzed in Ref.~\cite{Chen-Lou-ImpurityScatteringFriedel-2024}.

\begin{acknowledgments}
We acknowledge useful communications with M.~Amundsen and E.~Hodt. This work was supported by the Research Council of Norway through Grant No.~323766 and its Centres of Excellence funding scheme Grant No. 262633 ``QuSpin”. Support from Sigma2 - the National Infrastructure for High-Performance Computing and Data Storage in Norway, project NN9577K, is acknowledged.
\end{acknowledgments}

\appendix

\begin{widetext}
\section{Green's functions}
\label{sec:App-1}

\subsection{Rashba model}
\label{sec:App-1-Rashba}

The real-space Green's function in the Rashba model defined in Eq.~(\ref{Rashba-H}), see also Eq.~(\ref{Rashba-G-0-r}), at $\tilde{r}=0$ reads
\begin{equation}
\label{Rashba-G-r0}
G_{0}(\tilde{\omega},\mathbf{0},\mathbf{0}) =
-\nu_0 \frac{\left(\tilde{\omega}+1 +\tilde{h} \sigma_z\right) \ln{\left(\frac{|\tilde{k}_{+}|^2}{|\tilde{k}_{-}|^2}\right)}}{\tilde{k}^2_{+}-\tilde{k}_{-}^2} 
+\frac{\nu_0}{\tilde{k}^2_{+}-\tilde{k}_{-}^2} \sum_{\rho=\pm} \tilde{k}_{\rho}^2 \ln{\left(\frac{\tilde{\Lambda}^2}{|\tilde{k}_{\rho}|^2}\right)}
- i \pi \nu_0 \sum_{\rho =\pm} \frac{\left(\tilde{\omega}+1 -\tilde{k}_{\rho}^2 +\tilde{h} \sigma_z\right) \sign{\tilde{\omega}+1 -\tilde{k}_{\rho}^2} \Theta{(\tilde{k}_{\rho}^2)}}{\left|\tilde{k}^2_{+}-\tilde{k}_{-}^2\right|}, 
\end{equation}
where the momenta $\tilde{k}_{\rho}$ with $\rho=\pm$ are defined in Eq.~(\ref{Rashba-k}), $\tilde{\Lambda}$ is the momentum cutoff, and we consider only the energies such that at least one of the bands is intersected, i.e., $\tilde{\omega}>|\tilde{h}|-1$. In this case, $\tilde{k}_{\pm}$ are real. For $\tilde{\omega}<|\tilde{h}|-1$, Green's function is purely real and, hence, results in the vanishing LDOS, see Eq.~(\ref{model-nu-delta}).

The result at $\tilde{r}\neq0$ is more cumbersome,
\begin{eqnarray}
\label{Rashba-G-r}
G_{0}(\tilde{\omega},\tilde{\mathbf{r}},\mathbf{0}) &=& -\pi \nu_0 \sum_{\rho=\pm}\rho \frac{\left(\tilde{\omega}+1 -|\tilde{k}_{\rho}|^2 +\tilde{h} \sigma_z\right)Y_{0}{(|\tilde{k}_{\rho}|\tilde{r})}}{\tilde{k}^2_{+}-\tilde{k}_{-}^2}\nonumber\\
&-& i \pi \nu_0 \sum_{\rho=\pm} \frac{\left(\tilde{\omega}+1 -\tilde{k}_{\rho}^2 +\tilde{h} \sigma_z\right) \sign{\tilde{\omega}+1 -\tilde{k}_{\rho}^2} J_0{(|\tilde{k}_{\rho}|\tilde{r})} \Theta{(\tilde{k}_{\rho}^2)}}{\left|\tilde{k}^2_{+}-\tilde{k}_{-}^2\right|}  \nonumber\\
&+& i\sigma_y \pi \nu_0 \tilde{\alpha} e^{i\theta_r} \sum_{\rho=\pm} \rho \frac{|\tilde{k}_{\rho}|Y_{1}{(|\tilde{k}_{\rho}|\tilde{r})}}{\tilde{k}^2_{+}-\tilde{k}_{-}^2}  -\sigma_y \pi \nu_0 \tilde{\alpha} e^{i\theta_r} \sum_{\rho=\pm} \frac{|\tilde{k}_{\rho}|\sign{\tilde{\omega}+1 -\tilde{k}_{\rho}^2} J_1{(|\tilde{k}_{\rho}|\tilde{r})} \Theta{(\tilde{k}_{\rho}^2)}}{\left|\tilde{k}^2_{+}-\tilde{k}_{-}^2\right|},
\end{eqnarray}
where $J_n(x)$ is the Bessel function of the first kind, $Y_n(x)$ is the Bessel function of the second kind, and we assume that at least one band is filled, $\tilde{k}_{-}^2>0$. For $\tilde{k}_{\pm}^2<0$, one should replace  $Y_n(x) \to -2K_n(x)/\pi$ with $K_n(x)$ being the modified Bessel function of the second kind. In the above expressions, $\theta_r=0$ is used for $G_{0}(\tilde{\omega},\tilde{\mathbf{r}}, \mathbf{0})$ and $\theta_r=\pi$ for $G_{0}(\tilde{\omega},\mathbf{0}, \tilde{\mathbf{r}})$. Note that the phase cancels in $\tr{\, G_{0}(\tilde{\omega},\tilde{\mathbf{r}}, \mathbf{0}) T G_{0}(\tilde{\omega},\mathbf{0}, \tilde{\mathbf{r}})}$ and $\tr{\, \sigma_z G_{0}(\tilde{\omega},\tilde{\mathbf{r}}, \mathbf{0}) T G_{0}(\tilde{\omega},\mathbf{0}, \tilde{\mathbf{r}})}$.

In the derivation of Eqs.~(\ref{Rashba-G-r0}) and (\ref{Rashba-G-r}), we used the Jacobi–Anger expansion
\begin{equation}
\label{Rashba-Jacobi-Anger}
e^{ikr\cos{\theta}} = \sum_{m=-\infty}^{\infty} i^m J_m(kr) e^{im\theta},
\end{equation}
which allows one to straightforwardly integrate over the angle $\theta$ in Eq.~(\ref{Rashba-G-0-r}).

In addition, we use the standard asymptotic forms of the Bessel functions:
\begin{eqnarray}
\label{App-Rashba-J-rinf}
\lim_{x\to \infty} J_n(x) &\approx& \sqrt{\frac{2}{\pi x}} \cos{\left(x -\frac{n \pi}{2} -\frac{\pi}{4}\right)},\\
\label{App-Rashba-Y-rinf}
\lim_{x\to \infty} Y_n(x) &\approx& \sqrt{\frac{2}{\pi x}} \sin{\left(x -\frac{n \pi}{2} -\frac{\pi}{4}\right)}.
\end{eqnarray}

\subsection{\texorpdfstring{$p$-wave}{p-wave} magnets}
\label{sec:App-1-p-wave}

In this section, we present the inverse Fourier transform of the Green function (\ref{p-wave-2-G-0}) in the model of $p$-wave magnets defined in Eq.~(\ref{p-wave-2-H}). In calculating the integral over $\tilde{k}$, we use the Sokhotski–Plemelj formula, see Eq.~(\ref{Rashba-G-0-r}) for a similar integral in a Rashba model. The integrals can then be straightforwardly taken albeit are cumbersome.

For the sake of simplicity, we discuss different regimes separately.

We start with the case where $\tilde{k}^2_{\pm}$ contain an imaginary part, i.e., the expression under the square root in Eq.~(\ref{p-wave-2-a-k-pm}) is negative. This requires $\tilde{\omega}<-\tilde{J}-1$ and $\tilde{\alpha}^2>2\tilde{J}$; the latter is necessary to have local minima in the dispersion relation, see Fig.~\ref{fig:p-wave-2-eps}. We separate the real and imaginary parts in $\tilde{k}^2_{\pm}$ defined in Eq.~(\ref{p-wave-2-a-k-pm}) as
\begin{equation}
\label{p-wave-2-a-k-pm-1}
\tilde{k}_{\pm}^2 = \tilde{\omega} +1 +\frac{\tilde{\alpha}^2 \cos^2{\theta}}{2} \pm i\sqrt{\left|\tilde{\alpha}^2(1+\tilde{\omega}) \cos^2{\theta} +\tilde{J}^2 +\frac{\tilde{\alpha}^4\cos^4{\theta}}{4}\right|} \equiv K' \pm iK''.
\end{equation}

The inverse Fourier transform of Green's function at $\tilde{r}=0$ is
\begin{equation}
\label{p-wave-2-a-G-0-r0-int-1}
G_{\eta,0}(\tilde{\omega},\mathbf{0},\mathbf{0}) = \nu_0 \int_0^{2\pi} \frac{d\theta}{2\pi} \Bigg\{\frac{\tilde{\omega}+1 -K' +\eta \sigma_x \tilde{J}}{K''} \left[\arctan{\left(\frac{K'}{K''}\right)} +\arctan{\left(\frac{\tilde{\Lambda}^2}{K''}\right)}\right]
-\ln{\left(\frac{\Lambda^2}{\sqrt{(K')^2+(K'')^2}}\right)}
\Bigg\}.
\end{equation}
The Green function at $\tilde{r}\neq0$ reads
\begin{eqnarray}
\label{p-wave-2-a-G-0-rnot0-int-1}
&&G_{\eta,0}(\tilde{\omega},\tilde{\mathbf{r}},\mathbf{0}) =
\nu_0 \int_0^{2\pi} \frac{d\theta}{2\pi}
\sum_{\rho=\pm} \Bigg\{
i\sqrt{\pi}\frac{\tilde{\omega}+1 +K'+i\rho K'' +\rho \eta \sigma_x \tilde{J}}{2K''}
G_{13}^{21}{\left(-\frac{K'+i\rho K''}{4}\tilde{r}^2(\theta) \Bigg| \begin{array}{c}
0\\
0\, 0\, \frac{1}{2}
\end{array}\right)} \nonumber\\
&&+\pi \rho \sign{\tilde{r}(\theta)}\frac{\tilde{\omega}+1 -K'+i\rho K'' +\rho \eta \sigma_x \tilde{J}}{2K''} e^{i\sqrt{K'-i\rho K''} |\tilde{r}(\theta)|} \nonumber\\
&&+\sigma_z \frac{\sqrt{\pi} \tilde{\alpha} \cos{\theta}}{2K''}
\Big[
\sqrt{\pi} \sqrt{K' +i\rho K''} e^{i\sqrt{K' +i\rho K''} |\tilde{r}(\theta)|}
-i\sign{\tilde{r}(\theta)} \sqrt{K' +i\rho K''} G_{13}^{21}{\left(-\frac{K'+i\rho K''}{4} \tilde{r}^2(\theta) \Bigg| \begin{array}{c}
-\frac{1}{2}\\
-\frac{1}{2}\, \frac{1}{2}\, 0
\end{array}\right)}
\Big]
\Bigg\},\nonumber\\
\end{eqnarray}
where
\begin{equation}
\label{App-G-def}
G_{pq}^{mn}{\left(x \Bigg| \begin{array}{c}
a_1, \ldots q_p\\
b_1,\ldots b_q
\end{array}\right)} = \frac{1}{2\pi i} \int_{\gamma_L} dt\, x^{t} \frac{\prod_{j=1}^{m} \Gamma{\left(b_j -s\right)} \prod_{j=1}^{n} \Gamma{\left(1-a_j +s\right)}}{ \prod_{j=n+1}^{p} \Gamma{\left(a_j -s\right) \prod_{j=m+1}^{q} \Gamma{\left(1-b_j +s\right)}}}
\end{equation}
is the Meijer G-function~\cite{Gradshteyn-Ryzhik:book} and $\sqrt{-x} = -i\sqrt{|x|}$. The contour $\gamma_L$ is a loop beginning and ending at $+\infty$, that encircles the poles of the functions $\Gamma(b_j - s)$ once in the negative direction. All the poles of the functions $\Gamma(1- a_l +s)$ must remain outside this loop.

Next, we proceed to the case where both $\tilde{k}^2_{\pm}$ defined in Eq.~(\ref{p-wave-2-a-k-pm}) are real, i.e., the expression under the square root in Eq.~(\ref{p-wave-2-a-k-pm}) is positive, $\tilde{J}^2 +\tilde{\alpha}^4\cos^4{\theta}/4 +\tilde{\alpha}^2(1+\tilde{\omega}) \cos^2{\theta}>0$.
This inequality holds for all energies such that $\tilde{\omega}>-\tilde{J}-1$.

The Green function at $\tilde{r}=0$ reads
\begin{eqnarray}
\label{p-wave-2-a-G-0-r0-int}
G_{\eta,0}(\tilde{\omega},\mathbf{0},\mathbf{0}) &=& 2\nu_0 \int_0^{2\pi} \frac{d\theta}{2\pi} \int \tilde{k} d\tilde{k} G_{\eta, 0}(\tilde{\omega},\tilde{\mathbf{k}})
= - \nu_0 \int_0^{2\pi} \frac{d\theta}{2\pi} \Bigg\{\frac{1}{\tilde{k}_{+}^2 -\tilde{k}_{-}^2} \left[\left(\tilde{\omega}+1 +\eta \sigma_x\tilde{J}\right) \ln{\left|\frac{\tilde{k}_{+}^2}{\tilde{k}_{-}^2}\right|}
+\sum_{\rho=\pm} \rho \tilde{k}_{\rho}^2\ln{\left(\frac{\tilde{\Lambda}^2}{|\tilde{k}_{\rho}^2|}\right)} \right] \nonumber\\
&+& i\pi \sum_{\rho=\pm}\frac{\left(\tilde{\omega}+1 -\tilde{k}_{\rho}^2 +\eta \sigma_x\tilde{J}\right) \sign{\tilde{\omega}+1 -\tilde{k}_{\rho}^2}\Theta{(\tilde{k}_{\rho}^2)}}{\left|\tilde{k}_{+}^2 -\tilde{k}_{-}^2\right|}
\Bigg\}.
\end{eqnarray}

The inverse Fourier transform of the Green function is more involved at $\tilde{r}\neq0$:
\begin{eqnarray}
\label{p-wave-2-a-G-0-rnot0-int}
&&G_{\eta,0}(\tilde{\omega},\tilde{\mathbf{r}},\mathbf{0}) = 2\nu_0 \int_0^{2\pi} \frac{d\theta}{2\pi}\int \tilde{k} d\tilde{k}\, G_{\eta, 0}(\tilde{\omega},\tilde{\mathbf{k}}) e^{i\tilde{k}\tilde{r} \cos{(\theta_r -\theta)}}
= -\nu_0 \int_0^{2\pi} \frac{d\theta}{2\pi}
\Bigg\{-\sum_{\rho=\pm} \rho \frac{\tilde{\omega} +1 -\tilde{k}_{\rho}^2 +\eta \sigma_x \tilde{J}}{\tilde{k}_{+}^2 -\tilde{k}_{-}^2} \nonumber\\
&&\times \Bigg[i \pi \sign{\tilde{r}(\theta)} \left(\Theta{(\tilde{k}_{\rho}^2)} e^{-i \tilde{k}_{\rho} \tilde{r}(\theta)} +\Theta{(-\tilde{k}_{\rho}^2)} e^{-i \tilde{k}_{\rho} |\tilde{r}(\theta)|}\right) - e^{i \tilde{k}_{\rho} |\tilde{r}(\theta)|} \mbox{Ei}{\left(-i \tilde{k}_{\rho} |\tilde{r}(\theta)|\right)} -e^{-i \tilde{k}_{\rho} |\tilde{r}(\theta)|} \mbox{Ei}{\left(i \tilde{k}_{\rho} |\tilde{r}(\theta)|\right)} \Bigg] \nonumber\\
&&+i \pi \sum_{\rho=\pm}\frac{\left(\tilde{\omega} +1 -\tilde{k}_{\rho}^2 + \eta \sigma_x \tilde{J}\right) \sign{\tilde{\omega} +1 -\tilde{k}_{\rho}^2} e^{i\tilde{k}_{\rho} \tilde{r}(\theta)}}{\left|\tilde{k}_{+}^2 -\tilde{k}_{-}^2\right|} \nonumber\\
&&+ \sigma_z \tilde{\alpha} \cos{\theta}
\sum_{\rho=\pm} \frac{\rho \tilde{k}_{\rho} \sign{\tilde{r}(\theta)}}{\tilde{k}_{+}^2 -\tilde{k}_{-}^2} \Bigg[i \pi \left(\Theta{(\tilde{k}_{\rho}^2)} e^{-i \tilde{k}_{\rho} \tilde{r}(\theta)} +\sign{\tilde{r}(\theta)}\Theta{(-\tilde{k}_{\rho}^2)} e^{-i \tilde{k}_{\rho} |\tilde{r}(\theta)|}\right) \nonumber\\
&&+ e^{i \tilde{k}_{\rho} |\tilde{r}(\theta)|} \mbox{Ei}{\left(-i \tilde{k}_{\rho} |\tilde{r}(\theta)|\right)} -e^{-i \tilde{k}_{\rho} |\tilde{r}(\theta)|} \mbox{Ei}{\left(i \tilde{k}_{\rho} |\tilde{r}(\theta)|\right)} \Bigg]
+i \sigma_z \pi \tilde{\alpha} \cos{\theta} \sum_{\rho=\pm} \frac{ \tilde{k}_{\rho} \sign{\tilde{\omega} +1 -\tilde{k}_{\rho}^2} e^{i\tilde{k}_{\rho} \tilde{r}(\theta)}}{\left|\tilde{k}_{+}^2 -\tilde{k}_{-}^2\right|}
\Bigg\},
\end{eqnarray}

where $\tilde{r}(\theta) = \tilde{r} \cos{(\theta_r-\theta)}$ and $\mbox{Ei}(x) = -\int_{-x}^{\infty} dt\, e^{-t}/t$ is the exponential integral function.
\end{widetext}

\bibliography{p-wave-magnetis-Friedel-oscillations-v2}

\end{document}